\documentclass[pre, twocolumn]{revtex4-1}
\usepackage[dvipdfmx]{graphicx}
\usepackage{type1cm}
\usepackage[usenames,svgnames]{xcolor}
\usepackage{color}
\usepackage{url}
\usepackage{natbib}
\usepackage[colorlinks=true,linkcolor=blue,citecolor=blue,breaklinks=true]{hyperref}

\usepackage{amsmath,amssymb,bm,amsthm}
\usepackage{newtxtext,newtxmath}

\usepackage{braket}

\newcommand{\Tr}{\mathrm{Tr}\,}

\begin{document}
\title{Thermalization in open many-body systems based on eigenstate thermalization hypothesis}
\author{Tatsuhiko Shirai}
\email{tatsuhiko.shirai@aoni.waseda.jp}
\affiliation{
Green Computing Systems Research Organization, Waseda University, Tokyo 162-0042, Japan
}
\author{Takashi Mori}
\email{takashi.mori.fh@riken.jp}
\affiliation{
RIKEN Center for Emergent Matter Science (CEMS), Wako 351-0198, Japan
}

\begin{abstract}
We investigate steady states of macroscopic quantum systems under dissipation not obeying the detailed balance condition.
We argue that the Gibbs state at an effective temperature gives a good description of the steady state provided that the system Hamiltonian obeys the eigenstate thermalization hypothesis (ETH) and the perturbation theory in the weak system-environment coupling is valid in the thermodynamic limit.
We derive a criterion to guarantee the validity of the perturbation theory, which is satisfied in the thermodynamic limit for sufficiently weak dissipation when the Liouvillian is gapped for bulk-dissipated systems,
while the perturbation theory breaks down in boundary-dissipated chaotic systems due to the presence of diffusive transports.
We numerically confirm these theoretical predictions.
This paper suggests a connection between steady states of macroscopic open quantum systems and the ETH.
\end{abstract}

\maketitle

\section{Introduction}
A quantum system that is weakly coupled to a large environment usually relaxes to a steady state due to dissipation~\cite{Breuer_text,Weiss_text,Spohn1980}.
When the environment is in thermal equilibrium, the steady state is universally described by the Gibbs state by virtue of the detailed balance condition~\cite{Spohn-Lebowitz1978}.
In contrast, when the environment is out of equilibrium, the detailed balance condition is violated and there is no simple criterion to determine the steady state.
It is a challenge in statistical physics to predict the steady state in such a nonequilibrium situation~\cite{Mitra2006, Prosen2008, Prosen2011open, Prosen2011exact, Dhar2012, Sieberer2013, torre2013keldysh, Sieberer_review2016, Foss-Feig2017}.
Recent experimental progress using ultracold atoms and trapped ions has enabled us to introduce controlled dissipation~\cite{Barreiro2011,barontini2013,rauer2016,Tomita2017}, which leads us to the possibility of designing dissipation so that the steady state has desired properties~\cite{diehl2008quantum,verstraete2009quantum}.
This experimental background also motivates us to theoretically study steady states of quantum systems under dissipation not obeying the detailed balance condition.

In the present paper, the steady state of a quantum many-body system in a weak contact with an out-of-equilibrium environment is investigated.
It turns out that the Gibbs state at a certain effective temperature well describes the steady state in some open quantum systems despite the violation of the detailed balance condition.
We theoretically argue that there are two ingredients in the realization of a Gibbs steady state, i.e., the validity of the perturbation theory in weak dissipation and the eigenstate thermalization hypothesis (ETH), which is recognized as an important property of the Hamiltonian in explaining the approach to thermal equilibrium in \textit{isolated} quantum systems~\cite{Deutsch1991, Srednicki1994, Rigol2008, Mori2018}.

The ETH consists of two parts; the diagonal ETH and the off-diagonal ETH.
The diagonal ETH states that the diagonal elements of a local operator in the energy basis are smooth functions of the energy.
The off-diagonal ETH states that off-diagonal elements are irregularly fluctuating and exponentially small with respect to the system size.
The ETH is expected to hold in a wide class of nonintegrable Hamiltonians.

On the other hand, the validity of the weak-dissipation perturbation theory in macroscopic systems is highly nontrivial because it is known that its convergence radius quickly shrinks in the thermodynamic limit~\cite{Znidaric2015,Lemos2017}.
Here, we derive a theoretical criterion [Eq.~(\ref{eq:criterion}) below] of the validity of the perturbation theory by using the off-diagonal ETH, and then the diagonal ETH ensures that the steady state is well described by a Gibbs state if this criterion is satisfied.

In this way, this paper suggests a connection between steady states of macroscopic \textit{open} quantum systems and the ETH.
In a recent work~\cite{Ashida2018}, it was shown that a Gibbs state with a time-dependent temperature emerges in the transient dynamics of an open quantum system in which the system of interest is finite and obeys the ETH.
Our result should be distinguished from this recent result since we here focus on the steady state (i.e., the long-time limit) in a macroscopic open quantum system (i.e., the thermodynamic limit).

This paper is organized as follows.
Section~\ref{sec:review} gives a brief review of the weak-dissipation perturbation theory of the nonequilibrium steady state and the ETH.
In Sec~\ref{sec:theory}, we derive a theoretical criterion of the validity of the perturbation theory, which clarifies under what condition the ETH is relevant in characterizing the nonequilibrium steady state.
In Sec~\ref{sec:numerical}, we numerically confirm our theoretical prediction in a bulk-dissipated system and a boundary-dissipated system.
In Sec~\ref{sec:summary}, this paper is summarized with some future directions.

\section{Perturbative expansion and the ETH}\label{sec:review}
We consider a macroscopic system of the volume $V$ on $d$-dimensional cubic lattice that is in contact with an environmental system.
We denote by $H$ and $\rho$ the Hamiltonian and the reduced density matrix of the system of interest, respectively.
The dynamics of $\rho$ is assumed to be described by the Lindblad equation~\cite{Lindblad1976,Breuer_text}
\begin{equation}
\left\{
\begin{aligned}
\frac{d\rho}{dt}&=\mathcal{L}\rho=-i [H, \rho] +\gamma\mathcal{D}\rho, \\
\mathcal{D}\rho&=\sum_{a=1}^{V_{\rm D}} \left(L_{a}\rho L_{a}^{\dagger} -\frac{1}{2} \{ L_{a}^{\dagger} L_{a}, \rho\}\right),
\end{aligned}
\right.
\label{Lindblad}
\end{equation}
where $[\cdot,\cdot]$ and $\{\cdot,\cdot\}$ denote the commutator and the anti-commutator, respectively, and we put $\hbar=1$.
The dissipator $\mathcal{D}$ is characterized by the $V_{\rm D}$ Lindblad operators $\{L_a\}_{a=1,2,\dots,V_{\rm D}}$. When all the sites are subject to dissipation, $V_{\rm D}\propto V$, while dissipation acts only at the boundaries, $V_{\rm D}\propto V^{(d-1)/d}$.
The superoperator $\mathcal{L}$ is referred to as the Liouvillian.
Here, we consider the weak dissipation regime, i.e., small $\gamma$.

We assume that $H=\sum_nE_n\ket{n}\bra{n}$ obeys the ETH.
The ETH states that matrix elements $O_{nm}=\braket{n|O|m}$ of any local operator $O$ takes the following form with $\bar{E}_{nm}=(E_n+E_m)/2$ and $\omega_{nm}=E_n-E_m$:
\begin{equation}
O_{nm}=\braket{O}_{\beta_n}\delta_{nm}+\frac{r_{nm}}{\sqrt{D_{\bar{E}_{nm}}}}f_O(\bar{E}_{nm},\omega_{nm}),
\label{eq:ETH}
\end{equation}
where $f_O(\bar{E}_{nm},\omega_{nm})$ is a smooth function of the argument and decays exponentially fast for $|\omega_{nm}|\gtrsim\Lambda$ with a cutoff frequency $\Lambda$, which is independent of $V$.
The equilibrium expectation value of $O$ at the inverse temperature $\beta_n$ is denoted by $\braket{O}_{\beta_n}$, where $\beta_n$ is determined by the condition $E_n=\braket{H}_{\beta_n}$.
The quantity $D_E$ is the number of energy eigenstates with eigenvalues between $E-\Delta E$ and $E$ with some width $\Delta E$~\footnote{Since off-diagonal elements decay for $|E_n-E_m|\gtrsim\Lambda$, we should choose $\Delta E\sim \Lambda$. Otherwise, the function $f_O$ will diverge or vanish in the thermodynamic limit.}, and it scales as $D_E\sim e^{O(V)}$ with volume $V$.
The last term of Eq.~(\ref{eq:ETH}) expresses a small fluctuating part, and $\{r_{nm}\}$ behave as if their real and imaginary parts were random variables of mean zero and variance unity. 

The steady state $\rho_{\rm s}$ is defined by $\mathcal{L}\rho_{\rm s}=0$.
Since $\gamma$ is assumed to be small, we perform the perturbative expansion of $\rho_{\rm s}$ in $\gamma$:
\begin{equation}
\rho_{\rm s}=\sum_{n=0}^\infty\gamma^n\rho_{\rm s}^{(n)}.
\label{perturbation}
\end{equation}
By substituting this expression into $\mathcal{L}\rho_{\rm s}$ and requiring that it vanishes in each order in $\gamma$, we obtain
\begin{equation}
[H,\rho_{\rm s}^{(0)}]=0
\label{0th}
\end{equation}
for $O(\gamma^0)$, and
\begin{equation}
-i[H,\rho_{\rm s}^{(1)}]+\sum_{a} \left[ L_{a}\rho_{\rm s}^{(0)} L_{a}^{\dagger} -\frac{1}{2} \{ L_{a}^{\dagger} L_{a}, \rho_{\rm s}^{(0)} \} \right]=0
\label{1st}
\end{equation}
for $O(\gamma^1)$.
Equation~(\ref{0th}) implies that $\rho_{\rm s}^{(0)}$ is diagonal in the energy basis, i.e.
\begin{equation}
\rho_{\rm s}^{(0)}=\sum_n P_n \ket{n}\bra{n}.
\label{eq:rho_0}
\end{equation}
By looking at the $n$th diagonal element of Eq.~(\ref{1st}) in the energy basis, we obtain
\begin{equation}
\sum_m\left(W_{nm}P_m-W_{mn}P_n\right)=0,
\label{master}
\end{equation}
where $W_{nm}=\sum_a|\braket{n|L_a|m}|^2$~\cite{thingna2013reduced}.
Equation (\ref{master}) determines the diagonal elements of $\rho_{\rm s}^{(0)}$, i.e., $\{P_n\}$.
It is noted that $W_{nm}$ can be interpreted as the transition rate from the state $m$ to $n$.
The transition rates satisfy the detailed balance condition
\begin{equation}
\frac{W_{nm}}{W_{mn}}=e^{-\beta(E_n-E_m)}
\label{DBC}
\end{equation}
when the environment is in thermal equilibrium at the inverse temperature $\beta$~\cite{Davies_text,Davies1974,Spohn-Lebowitz1978}.
As a result, the steady state is given by the Gibbs state $P_n=e^{-\beta E_n}/Z(\beta)$ with the partition function $Z(\beta)=\sum_ne^{-\beta E_n}$.

When the environment is out of equilibrium, the detailed balance condition is violated, and hence $\{P_n\}$ is not necessarily of the Gibbs form.
Nevertheless, $\rho_{\rm s}^{(0)}$ is indistinguishable from the Gibbs state if the system Hamiltonian $H$ obeys the ETH.
Since $\beta_n$ in Eq.~(\ref{eq:ETH}) is almost constant, $\beta_n\approx\beta$, as long as the energy fluctuation in $\rho_{\rm s}^{(0)}$ is subextensive, we have
\begin{equation}
\Tr O\rho_{\rm s}^{(0)}\approx\braket{O}_\beta.
\end{equation}
In Appendix~\ref{app:temperature}, we show that the inverse effective temperature $\beta$ can be determined by numerically solving the following equation,
\begin{equation}
C(\beta)\equiv \sum_a \braket{[L_a^\dagger, H] L_a}_\beta=0.
\label{cbeta}
\end{equation}
Since $C(\beta)$ only depends on equilibrium values of local operators $[L_a^\dagger,H]L_a$, numerical methods for thermal equilibrium states, e.g. quantum Monte Carlo method, can be used to calculate $C(\beta)$.
In this way, the steady state is well described by the Gibbs state for small $\gamma$ as long as the naive perturbation theory is valid~\footnote{The effective temperature introduced here is operator independent in contrast to previous studies~\cite{Sieberer2013, sieberer2014nonequilibrium}}.

\section{Validity of perturbation theory}\label{sec:theory}
In Ref.~\cite{Lemos2017}, it is numerically shown that the convergence radius of the perturbative expansion of Eq.~(\ref{perturbation}) shrinks to zero in the thermodynamic limit, $V\to\infty$.
This means that it is a nontrivial issue whether the thermodynamic limit commutes with the weak-dissipation limit.
If they are commutable in evaluating the expectation value of an operator $O$, we have
\begin{equation}
\lim_{\gamma\to 0}\lim_{V\to\infty}\Tr O\rho_{\rm s}=\lim_{V\to\infty}\lim_{\gamma\to 0}\Tr O\rho_{\rm s}=\lim_{V\to\infty}\Tr O\rho_{\rm s}^{(0)}.
\label{limit}
\end{equation}
For macroscopic systems, the thermodynamic limit should be taken before the weak-dissipation limit, and hence, the left-hand side of Eq.~(\ref{limit}) is the quantity we want.
On the other hand, the right-hand side of Eq.~(\ref{limit}) corresponds to the solution in the leading-order perturbation theory.

In the present paper, we discuss whether Eq.~(\ref{limit}) holds by investigating the relative entropy density
\begin{equation}
s\equiv\frac{1}{V}S(\rho_{\rm s}\|\rho_{\rm s}^{(0)})=\frac{1}{V}\Tr\left[\rho_{\rm s}(\ln\rho_{\rm s}-\ln\rho_{\rm s}^{(0)})\right].
\label{dR}
\end{equation}
If $\lim_{\gamma\to 0}\lim_{V\to\infty}s=0$, we can conclude the \textit{macrostate equivalence} between $\rho_{\rm s}$ and $\rho_{\rm s}^{(0)}$~\cite{Touchette2015}, i.e., Eq.~(\ref{limit}) holds for intensive macroscopic observables $O$ that obey the large-deviation principle in the steady state~\cite{Mori2016}.

Now we derive a criterion for the validity of the perturbation theory, i.e. Eq.~(\ref{limit}).
We assume the open boundary condition for simplicity, but we can also derive the identical result for the periodic boundary condition.
The exact steady state $\rho_{\rm s}$ is decomposed as $\rho_{\rm s}=\rho_{\rm s}^{(0)}+\delta\rho$.
Then, the equality $\mathcal{L}\rho_{\rm s}=0$ is rewritten as
\begin{equation}
\delta\rho=-\tilde{\mathcal{L}}^{-1}\mathcal{L}\rho_{\rm s}^{(0)}=-\gamma\tilde{\mathcal{L}}^{-1}\mathcal{D}\rho_{\rm s}^{(0)},
\label{eq:delta_rho}
\end{equation}
where $\tilde{\mathcal{L}}^{-1}$ is the pseudo-inverse of $\mathcal{L}$.
From the definition of $\mathcal{D}$ and Eq.~(\ref{eq:rho_0}), we obtain
\begin{align}
\mathcal{D}\rho_{\rm s}^{(0)}=&\sum_{n\neq m}\sum_{a=1}^{V_{\rm D}}\left[\sum_l\braket{n|L_a|l}\braket{l|L_a^\dagger|m}P_l\right.
\nonumber \\
&\left.-\frac{1}{2}\braket{n|L_a^\dagger L_a|m}(P_n+P_m)\right]\ket{n}\bra{m}.
\end{align}
Since the Lindblad operators are assumed to be local, they obey the ETH [see Eq.~(\ref{eq:ETH})].
We consider an energy shell that consists of the energy eigenstates $\{\ket{n}\}$ with $E_n\in[\bar{E}-\delta E/2,\bar{E}+\delta E/2]$.
Here, $\bar{E}=\mathrm{Tr}\,H\rho_{\rm s}^{(0)}$ and $\delta E$ is chosen so that it is macroscopically small $\delta E=o(V)$ but large enough to ensure $\sum_{n: E_n\in[\bar{E}-\delta E/2,\bar{E}+\delta E/2]}P_n\approx 1$ (typically, $\delta E\sim V^{1/2}$)~\footnote{The difference between $\delta E\sim V^{1/2}$ here and $\Delta E\sim\Lambda$ in the ETH~(\ref{eq:ETH}) does not matter unless $\beta^{-1}\gg\Lambda$. For simplicity, below we do not take care of the difference between $\Delta E$ and $\delta E$. In more careful analysis, it turns out that $\beta$ in Eqs.~(\ref{eq:delta_M}), (\ref{eq:relative_criterion}), and (\ref{eq:criterion}) should be replaced by $\max\{\beta, 1/\Lambda\}$.}.
Then, each $P_n$ is roughly equal to $1/D$, where $D$ is the number of eigenstates within the energy shell.
We can then evaluate the order of magnitude of $\mathcal{D}\rho_{\rm s}^{(0)}$ as
\begin{equation}
\mathcal{D}\rho_{\rm s}^{(0)}\sim\sum_{n\neq m, |E_n-E_m|\lesssim\Lambda}\frac{V_{\rm D}^{1/2}R_{nm}}{D^{3/2}}\ket{n}\bra{m},
\label{eq:Drho}
\end{equation}
where $\{R_{nm}\}$ behave as random variables of mean zero and variance of $O(1)$, and $\Lambda$ is a cutoff frequency that is independent of $V$.

Next, we multiply $\tilde{\mathcal{L}}^{-1}$.
Let us define the gap $\gamma\Delta$ of the Liouvillian as the nonzero smallest absolute value of the eigenvalues of $\mathcal{L}$.
Then, the dominant contribution in multiplying $\tilde{\mathcal{L}}^{-1}$ comes from eigenmodes with eigenvalues close to $\gamma\Delta$, so we only consider such slow eigenmodes.
It is expected that matrix elements corresponding to fast oscillations, i.e. $|E_n-E_m|\gtrsim\gamma\Delta$, do not contribute to slow eigenmodes near the gap $\gamma\Delta$, and hence we can roughly evaluate the Frobenius norm of $\delta\rho$ by using Eqs.~(\ref{eq:delta_rho}) and (\ref{eq:Drho}) as follows:
\begin{align}
\|\delta\rho\|_F&:=\sqrt{\sum_{nm}|\braket{n|\delta\rho|m}|^2}
\nonumber\\
&\sim\frac{V_{\rm D}^{1/2}}{\Delta D^{3/2}}\sqrt{\sum_{n,m:|E_n-E_m|\lesssim\gamma\Delta}1}
\sim\sqrt{\frac{V_{\rm D}}{D}\frac{\beta\gamma}{\Delta}}.
\end{align}

Here, let us consider a macroscopic quantity $M=(1/V)\sum_{i=1}^VO_i$, where $i$ is an index of the lattice sites and $O_i$ is a local operator acting to sites near $i$.
In an energy shell, the diagonal elements of $M$ are roughly constant due to the ETH, so without loss of generality we put $\braket{M}_\beta=0$.
Then, the Frobenius norm of $M$ within the energy shell is evaluated as $\|M\|_F\approx\sqrt{D/V}$.
As a result, $\delta M:=\mathrm{Tr}\,M\delta\rho$ is evaluated as follows:
\begin{equation}
\left|\delta M\right|\leq\|M\|_F\|\delta\rho\|_F\sim\sqrt{\frac{V_{\rm D}}{V}\frac{\beta\gamma}{\Delta}}.
\label{eq:delta_M}
\end{equation}
Since the relative entropy density $s=S(\rho_{\rm s}\|\rho_{\rm s}^{(0)})/V$ is related to $\delta M$ by $ s \gtrsim |\delta M|^2$~\cite{Mori2016}, Eq.~(\ref{eq:delta_M}) implies
\begin{equation}
s\sim\frac{V_{\rm D}}{V}\frac{\beta\gamma}{\Delta}.
\label{eq:relative_criterion}
\end{equation}

Equation (\ref{eq:delta_M}) or Eq.~(\ref{eq:relative_criterion}) gives the following criterion for the validity of the perturbation theory:
\begin{equation}
\frac{V_{\rm D}}{V}\frac{\beta\gamma}{\Delta}\ll 1.
\label{eq:criterion}
\end{equation}
This is a main result of our paper.
In the case of bulk dissipation, i.e. $V_{\rm D}\propto V$, as long as the Liouvillian is gapped $\Delta>0$ in the thermodynamic limit, the criterion (\ref{eq:criterion}) is satisfied for a small but finite $\gamma$ in the thermodynamic limit.
The Liouvillian is expected to be gapped for a wide class of nonintegrable systems under bulk dissipation with no conserved quantity, and hence Eq.~(\ref{eq:criterion}) implies that the steady state is described by a Gibbs state for an equally wide class of open systems.
On the other hand, in the case of boundary dissipation, $V_{\rm D}\propto V^{(d-1)/d}$ and the criterion reads $\beta\gamma/(V^{1/d}\Delta)\ll 1$.
The Liouvillian gap in a boundary-dissipated chaotic system typically behaves as $\Delta\sim V^{-\theta/d}$ with an exponent $\theta\geq 1$~\cite{Znidaric2015}.
In many cases $\theta>1$, and then our theory predicts that the perturbation theory may break down for an arbitrarily small $\gamma$ in the thermodynamic limit~\footnote{One might expect $\theta=2$ for chaotic systems because of the presence of diffusive transport, but it is not always the case~\cite{Znidaric2015}.}.
Below, theoretical predictions discussed here will be numerically confirmed.

\section{Numerical result}\label{sec:numerical}
\begin{figure*}[t]
\centering
\includegraphics[width=0.49\textwidth]{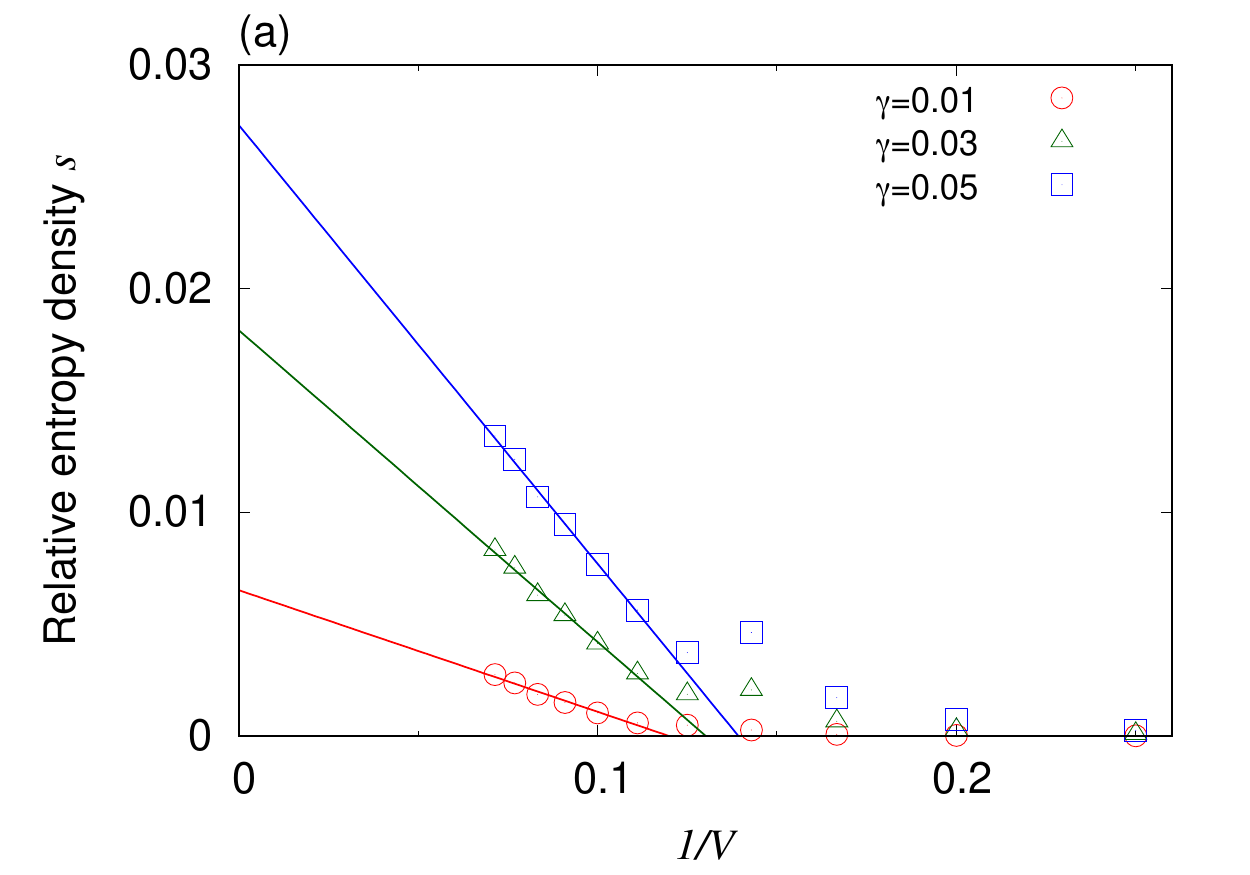}
\includegraphics[width=0.49\textwidth]{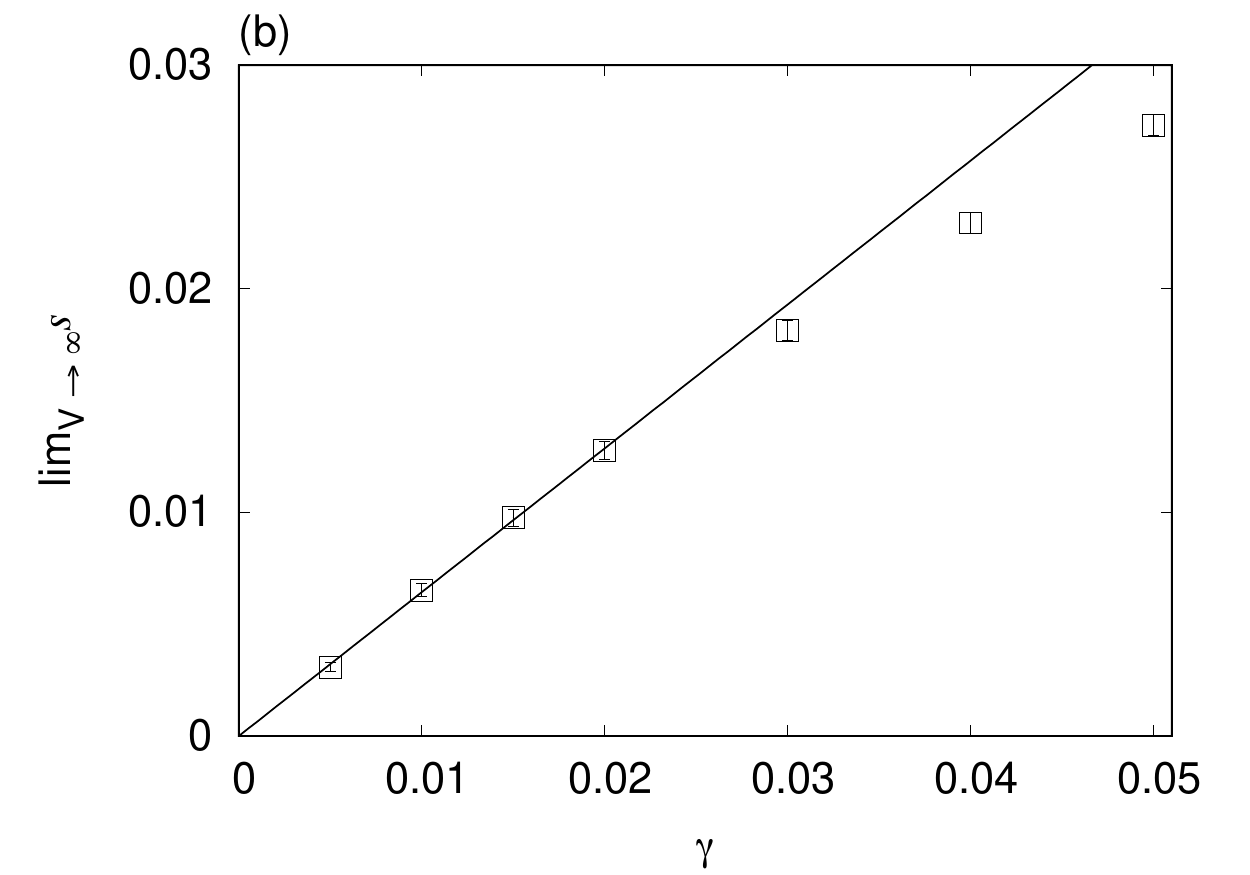}
\caption{
(a) System-size dependences of $s$ for the system with bulk dissipation [Eq.~(\ref{uniform})]: $\gamma=0.01$(circle), $0.03$(triangle), and $0.05$(square).
(b) $\gamma$-dependence of $\lim_{V\to\infty}s$, implying that $\lim_{\gamma \to 0}\lim_{V\to\infty}s=0$.
}
\label{NF}
\end{figure*}

\begin{figure*}[t]
\centering
\includegraphics[width=0.49\textwidth]{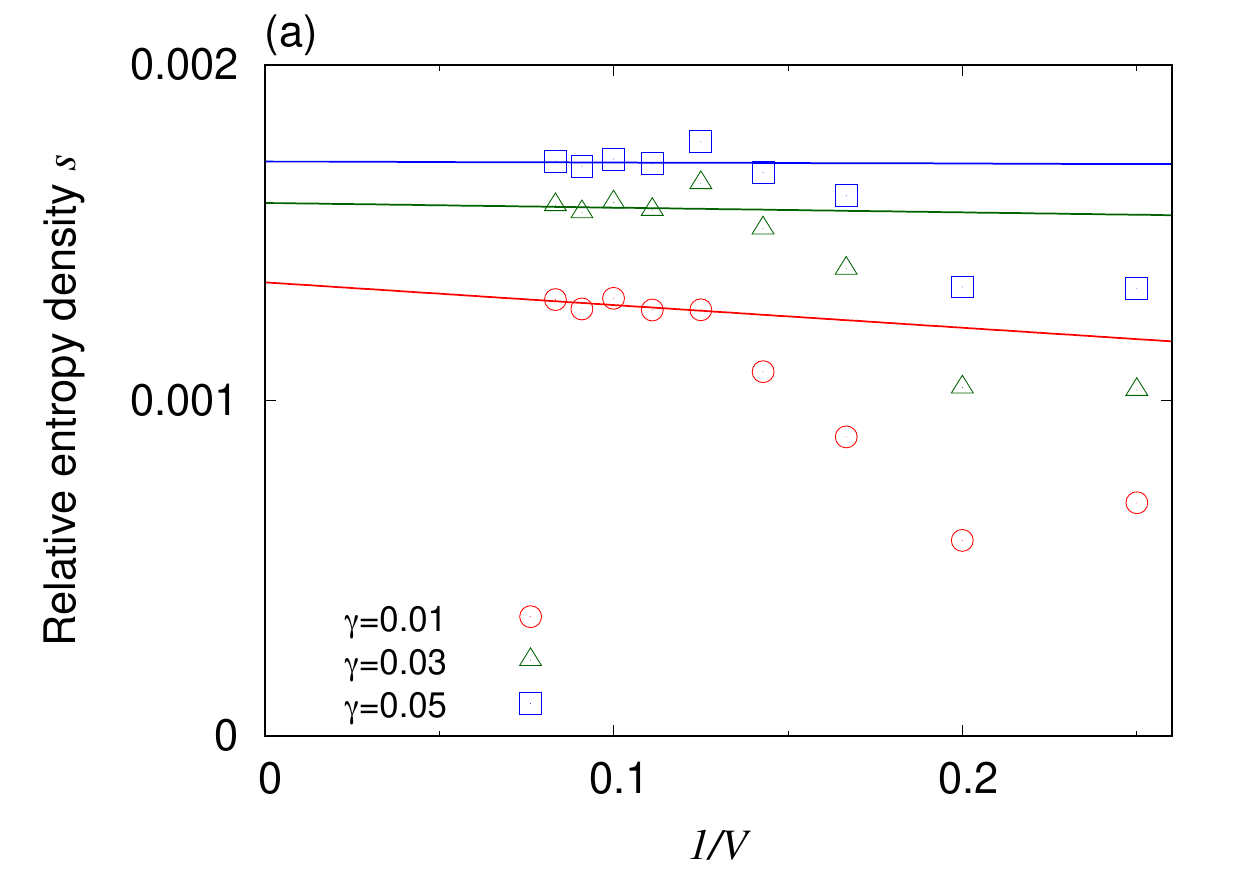}
\includegraphics[width=0.49\textwidth]{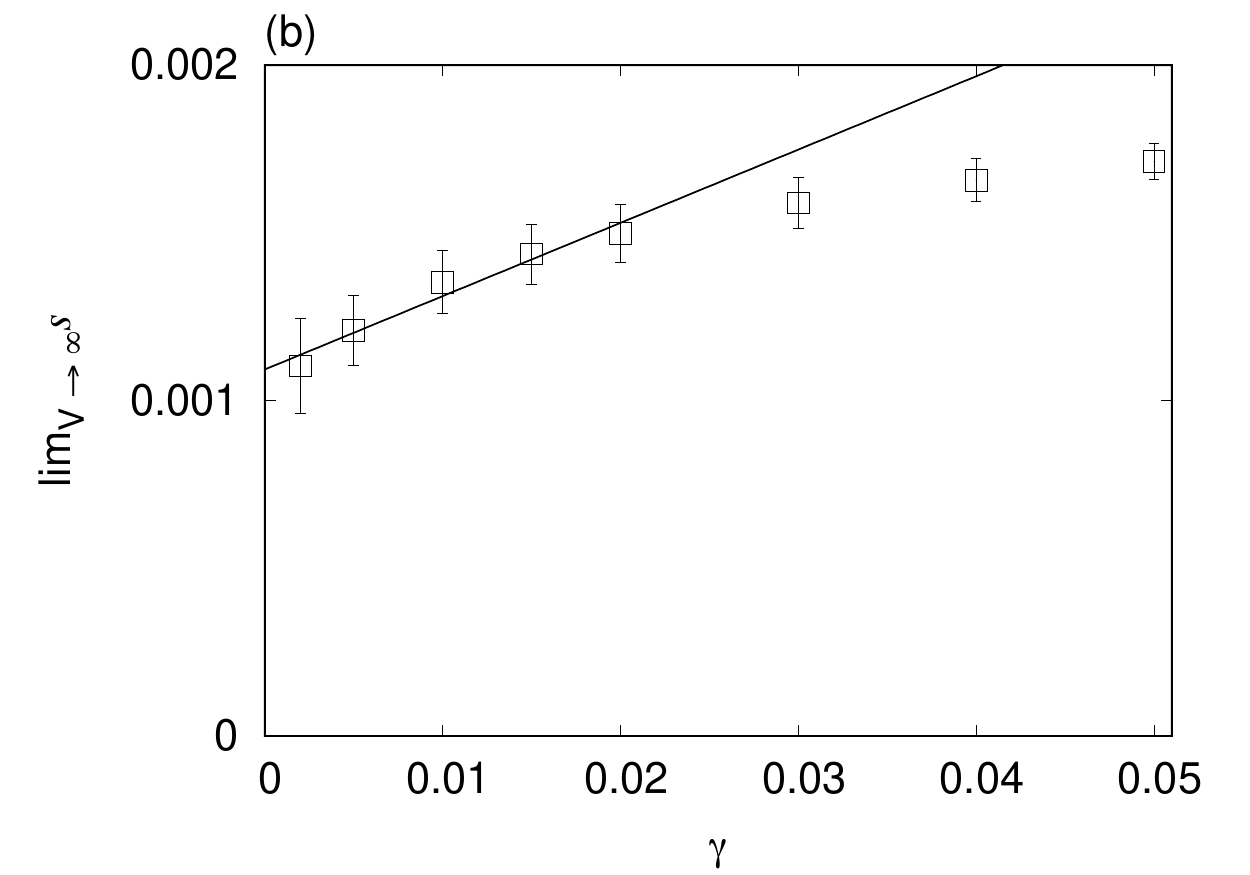}
\caption{
(a) System-size dependences of $s$ for the system with boundary dissipation [Eq.~(\ref{Bose_H}) and~(\ref{Bose_L})]: $\gamma=0.01$(circle), $0.03$(triangle), and $0.05$(square).
(b) $\gamma$-dependence of $\lim_{V\to\infty}s$, implying that $\lim_{\gamma \to 0}\lim_{V\to\infty}s \neq 0$.
}
\label{F}
\end{figure*}
\subsection{Bulk-dissipated system}
We first study a nonequilibrium system under bulk dissipation with no conserved quantity.
We consider the following dissipative Ising chain under the periodic boundary condition:
\begin{equation}
\left\{
\begin{aligned}
&H=\sum_{i=1}^V (h^z S_i^z + h^x S_i^x + g S_i^z S_{i+1}^z), \\
&L_i= S_i^- \quad (i=1,\cdots, V),
\end{aligned}
\right.
\label{uniform}
\end{equation}
where $\{ \vec{S}_i \}_{i=1}^V$ are spin-1/2 operators and $S_i^{\pm}\equiv S_i^x \pm iS_i^y$.
The parameters of the Hamiltonian are set as $(h^z,h^x,g)=(1.809, 1.618, 4)$, with which the ETH has been numerically shown to hold~\cite{Kim2014}.
This open quantum system has been implemented using Rydberg atoms~\cite{Carr2013, Letscher2017} and nonequilibrium phase transitions have been theoretically discussed~\cite{Lee2011}.
In this system the up and down spin states correspond to the Rydberg state and the ground state of an atom, respectively, and $\{ L_i \}$ describes the spontaneous emission in each atom.
It is noted that the detailed balance condition is not satisfied in this model.
In Appendix, the microscopic derivation of the Lindblad equation is given (Appendix~\ref{app:Lindblad}) and the violation of the detailed balance condition is demonstrated (Appendix~\ref{app:violation}).

The system-size dependences of $s$ at $\gamma=0.01, 0.03$ and $0.05$ are shown in Fig.~\ref{NF} (a).
We find the linear dependence of the distances on $1/V$ for large $V$ ($V\ge 9$).
By using this linear dependence, we extrapolate the data to the thermodynamic limit, and obtain $\lim_{V\to\infty}s$ for several small values of $\gamma$ [Fig.~\ref{NF} (b)].
We find $\lim_{V\to\infty}s\propto\gamma$ for $\gamma \lesssim 0.02$,
which is consistent with Eq.~(\ref{eq:relative_criterion}) with $V_{\rm D}=V$.
This conclusion is independent of the choice of the parameters $(h^z,h^x,g)$ as far as we have calculated.
In Appendix~\ref{app:another}, we provide another example showing the same $\gamma$-dependence of $s$.

\subsection{Boundary-dissipated system}

Next, we discuss a nonequilibrium system with boundary dissipation.
As we have already argued, our criterion~(\ref{eq:criterion}) tells us that the perturbation theory would break down and the steady state is not described by a Gibbs state in a boundary-dissipated system.
In order to understand this result more intuitively, suppose a one-dimensional system in contact with two particle reservoirs with different chemical potentials at each end.
The chemical potential difference drives the system, and particles will flow diffusively in the bulk.
Such diffusive transports result in a gradient in the particle density profile.
On the other hand, if the system Hamiltonian possesses translation invariance in the bulk, an individual energy eigenstate shows a uniform density profile, and hence its mixture like $\rho_{\rm s}^{(0)}$ cannot reproduce the expected gradient of the density profile in the steady state.
This argument can be generalized to other conserved currents (e.g., an energy current between two thermal reservoirs at different temperatures).
The perturbation theory fails in such a situation.

The failure of the perturbation theory is demonstrated for the hard-core Bose-Hubbard model driven by two environments with different chemical potentials:
\begin{equation}
\begin{aligned}
H=&\sum_{i=1}^{V-1} \left[-h (b_i b_{i+1}^{\dagger} + b_i^{\dagger} b_{i+1})+J\left(n_i-\frac{1}{2}\right)\left(n_{i+1}-\frac{1}{2}\right)\right]\\
&+ \sum_{i=1}^{V-2} \left[-h' (b_i b_{i+2}^{\dagger} + b_i^{\dagger} b_{i+2})+J' \left(n_i-\frac{1}{2}\right)\left(n_{i+2}-\frac{1}{2}\right)\right],
\end{aligned}
\label{Bose_H}
\end{equation}
and
\begin{equation}
\begin{aligned}
L_1=\sqrt{1+\mu}b_1^{\dagger}, \quad L_2=\sqrt{1-\mu}b_1, \\
L_3=\sqrt{1-\mu}b_V^{\dagger}, \quad L_4=\sqrt{1+\mu}b_V,
\end{aligned}
\label{Bose_L}
\end{equation}
where $b_i$ and $b_i^{\dagger}$ are annihilation and creation operators of a boson at site $i$, and $n_i=b_i^\dagger b_i= 0$ or $1$.
The parameters of the Hamiltonian are given by $(h, h', J, J')=(0.9167, 0.2449, 4, 0.9045)$.
The Lindblad operators $\{ L_{a} \}_{a=1}^4$ act on the boundaries of the lattice and $\mu$ effectively controls the chemical potential of the environments.
We set $\mu=0.1$.
In the steady state, we have a nonuniform particle density profile.

The system-size dependences of $s$ at $\gamma=0.01, 0.03$ and $0.05$ are shown in Fig.~\ref{F} (a).
Again, we find linear dependence on $1/V$ for $V \ge 9$, so the value in the thermodynamic limit is estimated by extrapolating the data.
In this way, we obtain the $\gamma$-dependence of $\lim_{V\to\infty}s$ [Fig.~\ref{F} (b)], showing that the distance is finite in the limit of $\gamma \to 0$: $\lim_{\gamma \to 0} \lim_{V \to \infty} s\neq 0$.
This result clearly shows the failure of the perturbation theory in a boundary-dissipated system.

\section{Summary}~\label{sec:summary}
In the present paper, we have investigated steady states of macroscopic quantum systems under dissipation not obeying the detailed balance condition.
We have theoretically argued that even in such nonequilibrium situations, the Gibbs state at effective temperature is a good description of the steady states.
There are two ingredients in emergence of the Gibbs state:
the validity of the weak-dissipation perturbation theory and the ETH.

We have derived a criterion for the validity of the perturbation theory beyond the convergence radius, which shrinks to zero in the thermodynamic limit~\cite{Znidaric2015, Lemos2017}.
It tells us that the perturbation theory works well for sufficiently weak bulk dissipation as long as the Liouvillian is gapped.
On the other hand, the perturbation theory breaks down for an arbitrarily weak boundary dissipation because of the vanishing gap of the Liouvillian due to the presence of diffusive transport.
Our numerical calculations have confirmed those theoretical predictions.

There remain some issues to be studied.
The effect of an extensive number of conserved quantities in integrable models on the steady states should be studied.
Our theoretical criterion (\ref{eq:criterion}) has been derived by using the off-diagonal ETH, which is not valid in integrable systems.
It is expected that the steady state is well described by a generalized Gibbs ensemble under a certain condition~\cite{Lenarcic2018}.
The extension of our theory to systems with finite dissipation strength are also important open problems.

\begin{acknowledgments}
The authors would like to thank Yuto Ashida, Ryusuke Hamazaki, Tomotaka Kuwahara and Keiji Saito for useful discussion. 
The numerical calculations have been done mainly on the supercomputer system at Institute for Solid State Physics, University of Tokyo.
This work was supported by the Japan Society for the Promotion of Science KAKENHI Grant No. JP18K13466 and No. JP19K14622.
\end{acknowledgments}

\appendix

\section{Convergence radius of perturbative series}\label{app:convergence}
\begin{figure*}[t]
\centering
\includegraphics[width=0.49\linewidth]{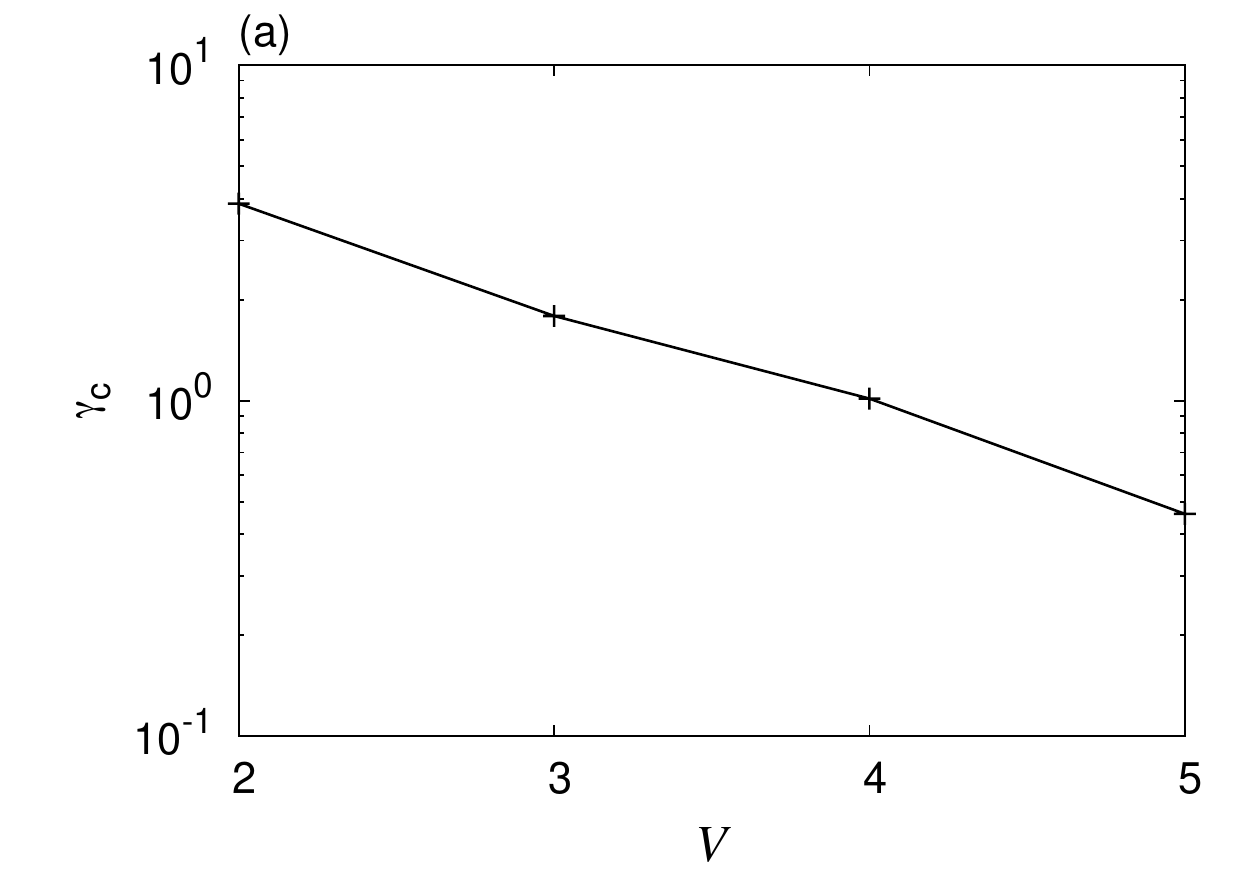}
\includegraphics[width=0.49\linewidth]{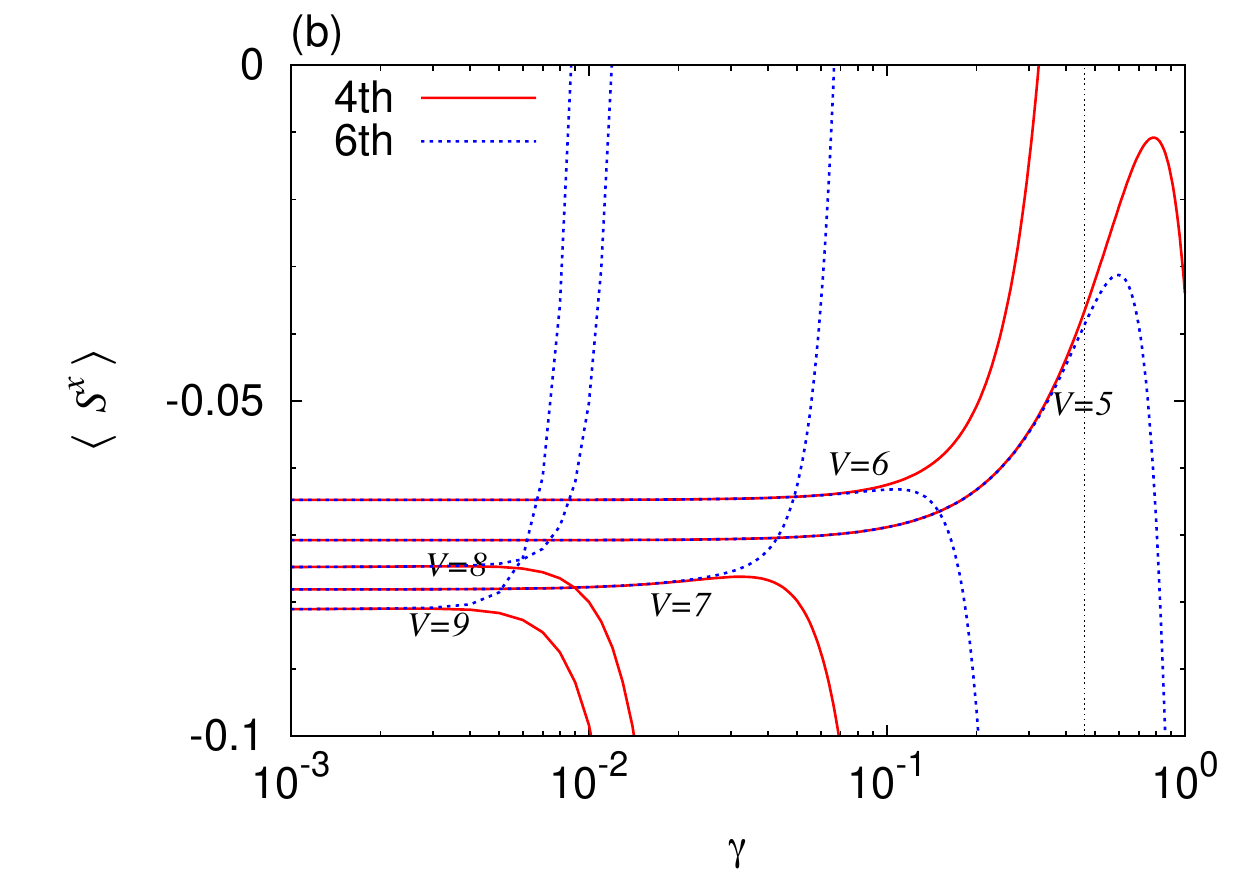}
\caption{
(a) System-size dependence of the convergence radius $\gamma_{\rm c}$.
(b) Expectation values of $\sum_{i=1}^V S_i^x/V$ over the fourth-order (bold) and the sixth-order (dotted) perturbative solution are presented for various system size, $5\le V \le 9$.
The vertical dotted line represents $\gamma_{\rm c}$ for $V=5$.
}
\label{radius}
\end{figure*}

Lemos and Prosen~\cite{Lemos2017} developed a numerical method to estimate the convergence radius of the perturbative expansion, $\gamma_{\rm c}$, and they argued that $\gamma_{\rm c}$ shrinks to zero with the system size for generic open quantum systems.
In the main text, we have performed the linear fitting of the numerical data for large system size at different values of $\gamma$ to obtain $\lim_{V\to\infty}s$.
In the argument, we have assumed that $\gamma=0.005$ is greater than $\gamma_{\rm c}$ for $V=9$. 
Here, we show that it is true for our model [Eq.~(\ref{uniform})].

In Fig.~\ref{radius}(a), the system-size dependence of $\gamma_{\rm c}$ is presented up to $V=5$.
The convergence radius $\gamma_{\rm c}$ shows the exponential decay with the system size, and it suggests that $\gamma_{\rm c}$ shrinks to zero in the thermodynamic limit.
However, as in Ref.~\cite{Lemos2017} we could not obtain $\gamma_{\rm c}$ for larger system size.

In order to estimate $\gamma_{\rm c}$ for larger system size $(V \ge 6)$, we compare two perturbative solutions, which are obtained by truncation of the perturbation series [Eq.~(\ref{perturbation})] up to fourth order and sixth order.
In Fig.~\ref{radius}(b), we plot the expectation values of $\sum_{i=1}^V S_i^x /V$ over each perturbative solution by solid curve (fourth order) and dotted curve (sixth order), respectively.
We found that two curves start to deviate at a certain value of $\gamma$, and the value is close to $\gamma_{\rm c}$ for $V=5$.
We use this relation to estimate the approximated values of $\gamma_{\rm c}$ for larger system size.
In Fig.~\ref{radius}(b), we found the approximated value of $\gamma_{\rm c}$ decreases with the system size and it is smaller than $0.005$ for $V=9$.

\section{Derivation of the Lindblad equation}\label{app:Lindblad}
We give a microscopic derivation of the Lindblad equation, Eq.~(\ref{uniform}).
The following derivation is essentially the same as the one found in Refs.~\cite{verso2010dissipation, torre2013keldysh} although the model is different.
Our microscopic model consists of a chain of laser-driven Rydberg atoms in contact with a bath of harmonic oscillators~\cite{Lee2011,Carr2013, Letscher2017}.
Each atom is regarded as a spin-1/2 (a two-level system).
The up and down spin states correspond to an excited Rydberg state and the ground state, respectively.
Let us denote by $h^x$ and $\omega$ the amplitude and the frequency of the laser driving.
Then, the Hamiltonian of the Rydberg atoms alone is given by
\begin{equation}
H_{\rm S}(t)=\sum_{i=1}^V \left((h^z+\omega)S_i^z+\frac{h^x}{2}(S_i^+ e^{-i\omega t}+S_i^- e^{i\omega t})+gS_i^zS_{i+1}^z\right),
\end{equation}
where $h^z$ is the detuning frequency and $g$ is the strength of interactions between neighboring atoms.

The Hamiltonian of the total system, including the bath, can be written as
\begin{equation}
H_{\rm T}(t)=H_{\rm S}(t)+\lambda H_{\rm I}+H_{\rm B},
\end{equation}
where $H_{\rm B}$ and $H_{\rm I}$ are the Hamiltonians of the bath and the interaction between the Rydberg atoms and the bath, respectively.
Here, $\lambda$ represents the interaction strength, which is assumed to be small.
The explicit forms of the Hamiltonians $H_{\rm B}$ and $H_{\rm I}$ are given by
\begin{equation}
\left\{
\begin{aligned}
H_{\rm B}=&\sum_{i=1}^V \sum_{\alpha} \omega_{\alpha}a_{i,\alpha}^{\dagger}a_{i,\alpha},\\
H_{\rm I}=&\sum_{i=1}^V \sum_{\alpha} (S_i^+ a_{i,\alpha} +S_i^- a_{i,\alpha}^{\dagger}),
\end{aligned}
\right.
\end{equation}
where $a_{i,\alpha}$ and $a_{i,\alpha}^\dagger$ are bosonic annihilation and creation operators of the bath modes, respectively.
It is assumed that each spin is coupled to its own thermal bath independently.

The time dependence of the Hamiltonian can be eliminated by moving to a rotating frame.
For this purpose, we introduce a unitary operator,
\begin{equation}
U(t)=e^{-i\omega t \sum_{i=1}^V (S_i^z+\sum_\alpha a_{i,\alpha}^\dagger a_{i,\alpha})}.
\end{equation}
The total Hamilton in the rotating frame reads
\begin{equation}
H_{\rm T}^{\rm R}=U^{\dagger}(t) \left( H_{\rm T}(t)-i\frac{\partial}{\partial t} \right)U(t)=H+\lambda H_{\rm I}+H_{\rm B}^{\rm R},
\label{HTR}
\end{equation}
where $H$ and $H_{\rm B}^{\rm R}$ are given by Eq.~(\ref{uniform}) and
\begin{equation}
H_{\rm B}^{\rm R}=\sum_{i=1}^V \sum_\alpha (\omega_{\alpha}-\omega) a_{i,\alpha}^{\dagger}a_{i,\alpha},
\end{equation}
respectively.
Note that the energy of each bath mode in $H_{\rm B}^{\rm R}$ is shifted by $\omega$ with respect to that in $H_{\rm B}$.
Then, the bath in the rotating frame does not satisfy the Kubo-Martin-Schwinger (KMS) relation [see Eq.~(\ref{violation_KMS}) below], which results in the violation of the detailed balance condition [see Appendix~\ref{app:violation}].
Therefore, in the rotating frame, the problem is equivalent to that of an open quantum system in contact with an out-of-equilibrium environment.

The relaxation dynamics of the system of interest with a weak system-bath coupling is described by a Markovian quantum master equation.
By applying the Born-Markov approximation, we obtain the following master equation~\cite{kubo1991statistical, Breuer_text}, 
\begin{align}
\frac{d\rho}{dt}=&-i[H,\rho]+\lambda^2\sum_{i=1}^V\int_{-\infty}^{\infty} d\epsilon \lim_{\delta\to +0}\int_0^{\infty} du e^{-\delta u} \nonumber\\
&\times \left[([S_i^-, \rho S_i^+(-u)]e^{-i\epsilon u}-[S_i^+, S_i^-(-u) \rho]e^{i\epsilon u})G_1^{\rm R}(\epsilon) \right.\nonumber\\
&\left.+([S_i^+, \rho S_i^-(-u)]e^{-i\epsilon u}-[S_i^-, S_i^+(-u) \rho]e^{i\epsilon u})G_2^{\rm R}(\epsilon)\right],
\label{BornMarkov}
\end{align}
where $\rho$ is the density matrix of the system of interest in the rotating frame and $S_i^{\pm}(u)=e^{iHu}S_i^{\pm}e^{-iHu}$.
Here, $G_1^{\rm R}(\epsilon)$ and $G_2^{\rm R}(\epsilon)$ denote the bath correlation functions in the rotating frame,
\begin{equation}
\left\{
\begin{aligned}
G_1^{\rm R}(\epsilon)&\equiv \sum_\alpha \int_{-\infty}^{\infty} {\rm Tr} ( e^{iH_{\rm B}^{\rm R} t}a_{i,\alpha} e^{-iH_{\rm B}^{\rm R} t} a_{i,\alpha}^{\dagger}\rho_{\rm B}) e^{-i\epsilon t}\frac{dt}{2\pi},\\
G_2^{\rm R}(\epsilon)&\equiv \sum_\alpha \int_{-\infty}^{\infty} {\rm Tr} ( e^{iH_{\rm B}^{\rm R} t}a_{i,\alpha}^\dagger e^{-iH_{\rm B}^{\rm R} t} a_{i,\alpha} \rho_{\rm B}) e^{-i\epsilon t}\frac{dt}{2\pi},
\end{aligned}
\right.
\end{equation}
where $\rho_{\rm B}$ is a Gibbs state at inverse bath temperature $\beta_{\rm B}$: $\rho_{\rm B}=e^{-\beta_{\rm B} H_{\rm B}}/{\rm Tr} e^{-\beta_{\rm B} H_{\rm B}}$.
It is noted that the Gibbs state is invariant under the unitary transformation, i.e. $U(t) \rho_{\rm B} U^{\dagger}(t)=\rho_{\rm B}$.
The bath correlation functions in the rotating frame are related to those in the original frame $G_1(\epsilon)$ and $G_2 (\epsilon)$ as
\begin{equation}
G_1^{\rm R}(\epsilon)=G_1(\epsilon - \omega) \text{ and } G_2^{\rm R}(\epsilon)=G_2(\epsilon + \omega),
\end{equation}
where
\begin{equation}
\left\{
\begin{aligned}
G_1(\epsilon)&\equiv \sum_\alpha \int_{-\infty}^{\infty} {\rm Tr} ( e^{iH_B t}a_{i,\alpha} e^{-iH_B t} a_{i,\alpha}^{\dagger}\rho_B) e^{-i\epsilon t}\frac{dt}{2\pi},\\
G_2(\epsilon)&\equiv \sum_\alpha \int_{-\infty}^{\infty} {\rm Tr} ( e^{iH_B t}a_{i,\alpha}^\dagger e^{-iH_B t} a_{i,\alpha} \rho_B) e^{-i\epsilon t}\frac{dt}{2\pi}.
\end{aligned}
\right.
\end{equation}
The correlation functions in the original frame satisfy the KMS relation~\cite{kubo1957statistical},
\begin{equation}
G_1(\epsilon)=G_2(-\epsilon)e^{\beta_B \epsilon},
\label{KMS}
\end{equation}
while those in the rotating frame do not satisfy this relation;
\begin{align}
G_1^{\rm R}(\epsilon)&=G_1(\epsilon - \omega)\nonumber\\
&=G_2(-\epsilon + \omega)e^{\beta_{\rm B} (\epsilon - \omega)}\nonumber\\
&=G_2^{\rm R}(-\epsilon) e^{\beta_{\rm B} (\epsilon - \omega)} \neq G_2^{\rm R}(-\epsilon) e^{\beta_{\rm B} \epsilon}.
\label{violation_KMS}
\end{align}
As is mentioned before, this implies that the bath is in thermal equilibrium in the original frame but not in the rotating frame.

The integration over $u$ in Eq.~(\ref{BornMarkov}) gives
\begin{widetext}
\begin{align}
\frac{d\rho}{dt}=-i[H,\rho]+\lambda^2\sum_{i=1}^V \sum_{n,m}\int_{-\infty}^{\infty} d\epsilon \left[\left([S_i^-, \rho \ket{n}\bra{n} S_i^+ \ket{m}\bra{m}]\left(\pi \delta(\epsilon+E_n-E_m) -i {\rm P}\left(\frac{1}{\epsilon+E_n-E_m}\right)\right)\right.\right.\nonumber\\
\left.-[S_i^+, \ket{m}\bra{m}S_i^-\ket{n}\bra{n} \rho]\left(\pi \delta(\epsilon+E_n-E_m) +i {\rm P}\left(\frac{1}{\epsilon+E_n-E_m}\right)\right)\right)G_1(\epsilon-\omega)\nonumber\\
+\left([S_i^+, \rho \ket{n}\bra{n} S_i^- \ket{m}\bra{m}]\left(\pi \delta(\epsilon+E_n-E_m) -i {\rm P}\left(\frac{1}{\epsilon+E_n-E_m}\right)\right)\right.\nonumber\\
\left.\left.-[S_i^-, \ket{m}\bra{m} S_i^+ \ket{n}\bra{n} \rho]\left(\pi \delta(\epsilon+E_n-E_m) +i {\rm P}\left(\frac{1}{\epsilon+E_n-E_m}\right)\right) \right)G_2(\epsilon+\omega)\right],
\label{BornMarkov2}
\end{align}
\end{widetext}
where $H\ket{n}=E_n \ket{n}$ and ${\rm P}$ denotes the Cauchy principal value.
Let us denote by $D[\rho]$ the dissipation part of Eq.~(\ref{BornMarkov2}).
Then, Eq.~(\ref{BornMarkov2}) is written as $d\rho/dt=-i[H,\rho]+ D[\rho]$.

The master equation, Eq.~(\ref{BornMarkov2}), is simplified under the assumptions that $\omega \gg \max\{h^z,h^x,g\}$.
It is proved that matrix elements $\braket{n|S_i^{\pm}|m}$ decay exponentially in $|E_n-E_m|$, and the dominant contribution in the integral over $\epsilon$ in Eq.~(\ref{BornMarkov2}) comes from $|\epsilon|\lesssim\max\{h^z,h^x,g\}\ll\omega$.
Therefore, we can neglect the dependences of correlation functions on $\epsilon$,
\begin{equation}
G_1(\epsilon-\omega)\simeq G_1(-\omega) \text{ and } G_2(\epsilon+\omega)\simeq G_2(\omega).
\end{equation}
In this approximation, the Cauchy principal integrals become zero, and thus the dissipation part of Eq.~(\ref{BornMarkov2}) reads
\begin{align}
D[\rho]\simeq&2\pi \lambda^2 G_1(-\omega) \sum_{i=1}^V \left[ S_i^- \rho S_i^+ -\frac{1}{2}\{S_i^+ S_i^-, \rho\}\right.\nonumber\\
&\qquad\qquad \left.+e^{-\beta_{\rm B} \omega} \left( S_i^+ \rho S_i^- -\frac{1}{2}\{S_i^- S_i^+, \rho\} \right)\right],
\end{align}
where we have used the KMS condition, Eq.~(\ref{KMS}).
If we further assume that $\omega \gg 1/\beta_{\rm B}$, we obtain
\begin{equation}
D[\rho]\simeq \gamma \sum_{i=1}^V \left( S_i^- \rho S_i^+ -\frac{1}{2}\{S_i^+ S_i^-, \rho\} \right),
\end{equation}
where $\gamma\equiv 2\pi \lambda^2 G_1(-\omega)$.
This is identical to the Lindblad form in main text, Eq.~(\ref{uniform}).
The Lindblad operators $\{ S_i^- \}$ describe the transition from the up spin state to the down spin state due to dissipation.

\section{Effective temperature}\label{app:temperature}
\begin{figure*}[t]
\centering
\begin{tabular}{cc}
\includegraphics[width=0.49\linewidth]{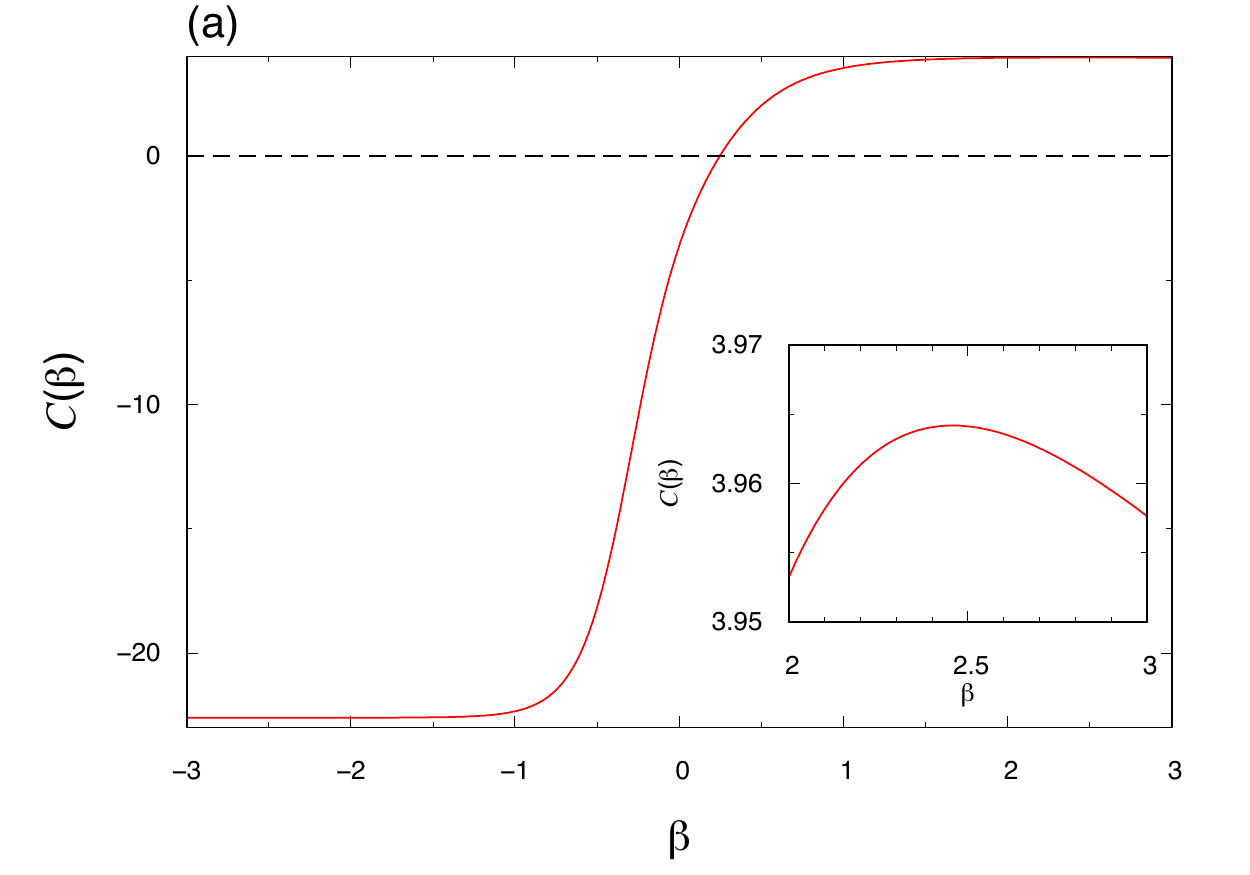}&
\includegraphics[width=0.49\linewidth]{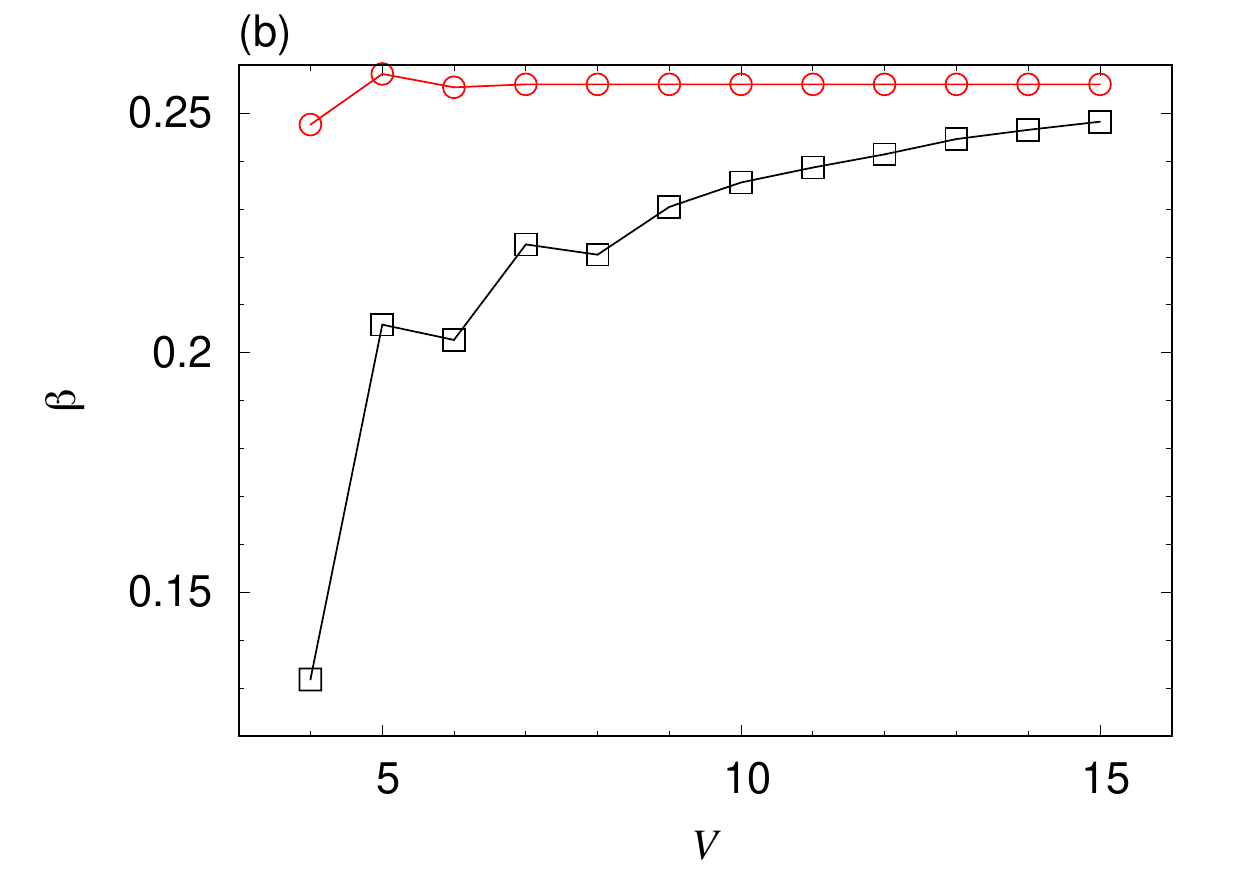}
\end{tabular}
\caption{
(a) $C(\beta)$ for system size $V=4$.
The crossing point of $C(\beta)$ (red solid curve) and zero (black dotted line) marks the inverse effective temperature $\beta$. Inset: Magnification of $C(\beta)$ around a maximum $(\beta\simeq 2.46)$.
(b) System-size dependences of inverse effective temperature $\beta$, which are obtained by two different methods: $C(\beta)=0$ (red circles) and ${\rm Tr} H \rho_{\rm s}^{(0)}={\rm Tr} H \rho_{\beta}$ (black squares).
}
\label{cbeta_N04}
\end{figure*}

We have shown in the main text that the steady state is locally indistinguishable from the Gibbs state $\rho_\beta =e^{-\beta H}/\Tr e^{-\beta H}$ for some open quantum systems that obey the Lindblad equation
\begin{equation}
\frac{d\rho}{dt}=-i[H,\rho]+\gamma \sum_a\left(L_a \rho L_a^{\dagger}-\frac{1}{2}\{L_a^\dagger L_a,\rho\}\right).
\label{Lindblad_supple}
\end{equation}
Here, $\beta$ is an inverse effective temperature that characterizes the steady state.

We provide a method to estimate $\beta$.
The steady state of the Lindblad equation $\rho_{\rm s}$ is obtained by setting the right-hand side of Eq.~(\ref{Lindblad_supple}) to zero.
By multiplying it by $H$ from the left and then taking its trace, we obtain the following equation:
\begin{equation}
\mathrm{Re}\sum_a\Tr[L_a^\dagger,H]L_a\rho_{\rm s}=0.
\label{energy_supple}
\end{equation}
Since $[L_a^\dagger,H]L_a$ is a local operator, $\rho_{\rm s}$ in Eq.~(\ref{energy_supple}) can be replaced by $\rho_\beta$. 
Then, we obtain Eq.~(\ref{cbeta}) in the main text, i.e.,
\begin{equation}
C(\beta)\equiv \sum_a {\rm Tr} ([L_a^\dagger, H] L_a \rho_\beta)=0,
\label{C_beta}
\end{equation}
where $\mathrm{Re}$ has been omitted since $\Tr[L_a^\dagger,H]L_a\rho_\beta$ is always real.
We can estimate the value of $\beta$ by using Eq.~(\ref{C_beta}).

There exists at least one solution because $C(\beta)$ is a continuous function of $\beta$ and satisfies $C(+\infty)\geq 0$ and $C(-\infty)\leq 0$.
We can prove $C(+\infty)\geq 0$ as follows.
At $\beta=+\infty$, $\rho_{\beta}=\ket{\Psi_{\rm min}}\bra{\Psi_{\rm min}}$ where $\ket{\Psi_{\rm min}}$ is the ground state of $H$ with energy $E_{\rm min}$.
Then,
\begin{align}
C(+\infty)&=\sum_a\bra{\Psi_\mathrm{min}} [L_a^\dagger,H]L_a\ket{\Psi_\mathrm{min}},
\nonumber \\
&=\sum_a\bra{\Psi_\mathrm{min}}L_a^\dagger HL_a\ket{\Psi_\mathrm{min}} -E_\mathrm{min}\bra{\Psi_\mathrm{min}}L_a^\dagger L_a\ket{\Psi_\mathrm{min}},
\nonumber \\
&=\sum_a\bra{\Psi_\mathrm{min}}L_a^\dagger L_a\ket{\Psi_\mathrm{min}}\left(\bra{\Phi_a}H\ket{\Phi_a}-E_\mathrm{min}\right),
\nonumber \\
&\geq 0,
\end{align}
where
\begin{equation}
\ket{\Phi_a}=\frac{L_a\ket{\Psi_\mathrm{min}}}{\sqrt{\bra{\Psi_\mathrm{min}}L_a^\dagger L_a\ket{\Psi_\mathrm{min}}}}.
\end{equation}
We can also prove $C(-\infty)\leq 0$ in the similar way.

We apply this method to the Lindblad equation in main text [see Eq.~(\ref{uniform})].
In Fig.~\ref{cbeta_N04} (a), we plot $C(\beta)$ for the system size $V=4$.
As it is mentioned above, $C(\beta)$ is negative at $\beta=-\infty$ $[C(-\infty)\simeq-22.6]$ and positive at $\beta=+\infty$ $[C(+\infty)\simeq3.83]$.
Between them, $C(\beta)$ has a maximum around $\beta \simeq 2.46$, and thus it is not a monotonic function of $\beta$.
In the present case, there is only one solution for $C(\beta)=0$, that determines the inverse effective temperature $\beta$.
In Fig.~\ref{cbeta_N04} (b), the system-size dependence of $\beta$ is depicted by red circles.
The estimated values of $\beta$ show a weak system-size dependence, and they are almost converged to $\beta \simeq 0.256$ for $V \geq 7$.

The inverse effective temperature $\beta$ is also evaluated by comparing the expectation values of the energy between $\rho_{\rm s}^{(0)}$ and $\rho_\beta$:
\begin{equation}
{\rm Tr} H \rho_{\rm s}^{(0)} = {\rm Tr} H \rho_\beta.
\label{direct}
\end{equation}
In Fig.~\ref{cbeta_N04} (b), we also plot the system-size dependence of $\beta$ obtained in this method by black squares.
The estimated values of $\beta$ in two methods approach each other with increasing system size.
It seems that $\beta$ obtained by $C(\beta)=0$ quickly converges to the value in the thermodynamic limit.
It should be emphasized that the estimation of $\beta$ using Eq.~(\ref{direct}) is a numerically hard task since it requires the evaluation of the transition rates $W_{ij}$ for all the pairs of energy eigenstates (to do so, we have to diagonalize the Hamiltonian), while solving Eq.~(\ref{C_beta}) is much easier since it only requires the calculation of the equilibrium expectation values of $[L_a^\dagger,H]L_a$.

\section{Violation of the detailed balance condition}\label{app:violation}
\begin{figure*}[t]
\centering
\begin{tabular}{cc}
\includegraphics[width=0.49\linewidth]{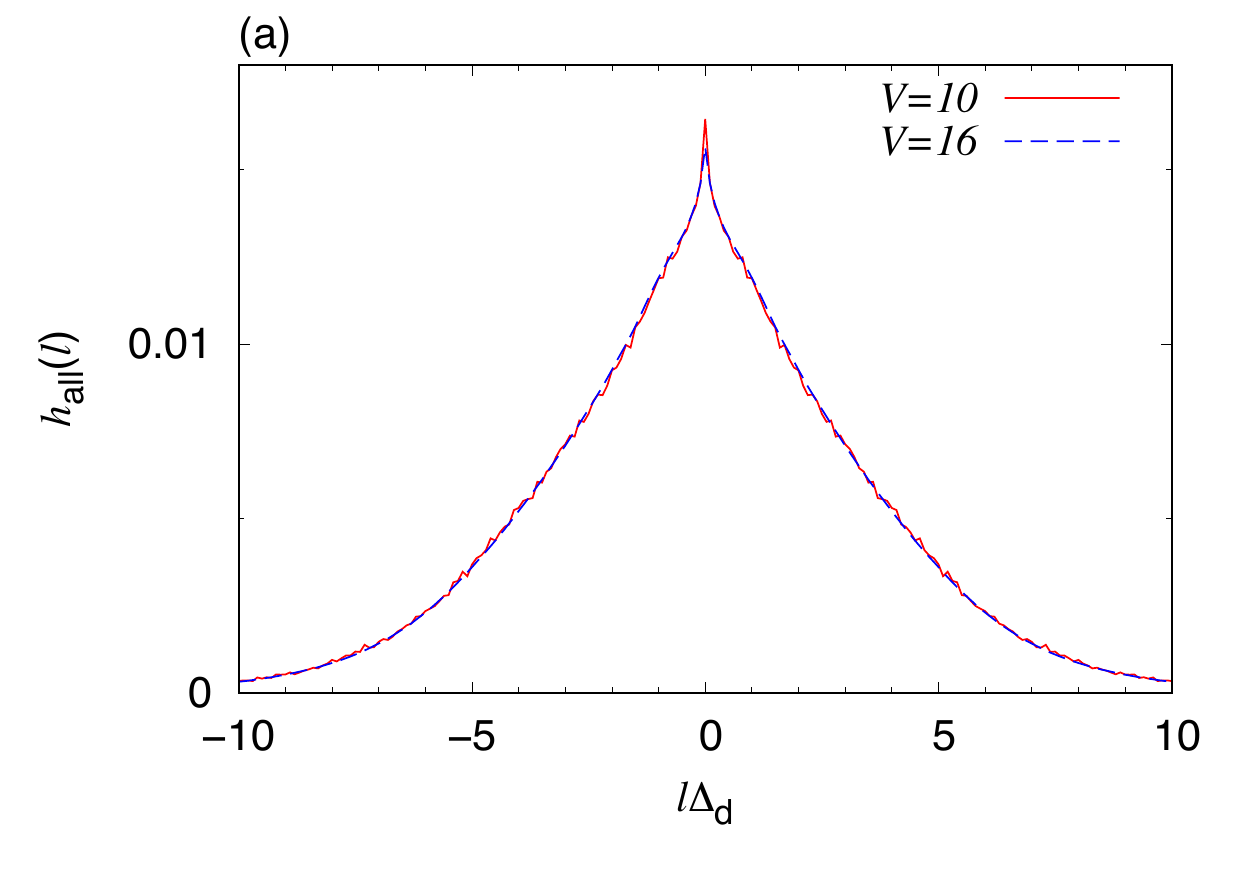}
\includegraphics[width=0.49\linewidth]{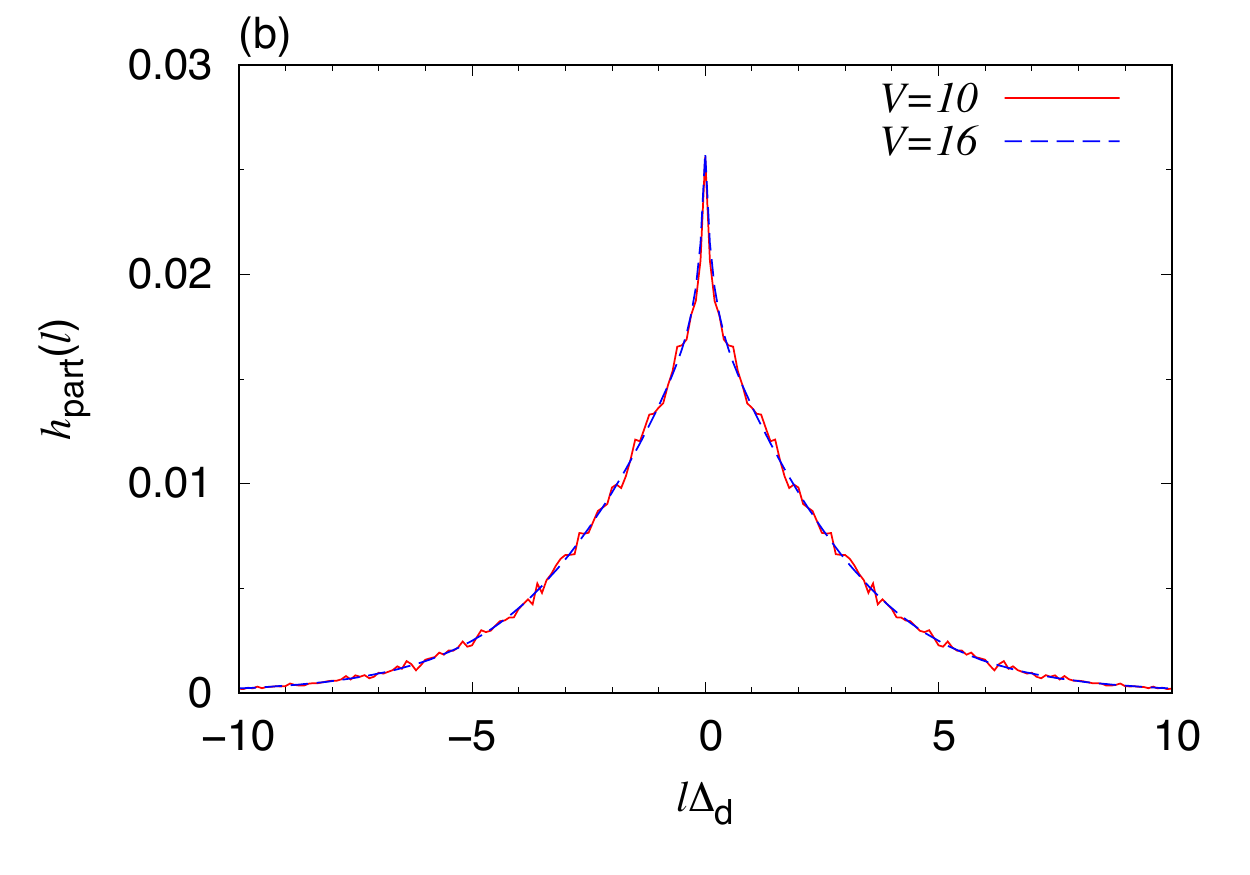}
\end{tabular}
\caption{
The scaled histograms for different system sizes, $V=10$ (red solid line) and $V=16$ (blue dotted line): (a) $h_{\rm all}(\ell)$ and (b) $h_{\rm part}(\ell)$.
}
\label{DB}
\end{figure*}

\begin{figure*}[t]
\centering
\begin{tabular}{cc}
\includegraphics[width=0.5\linewidth]{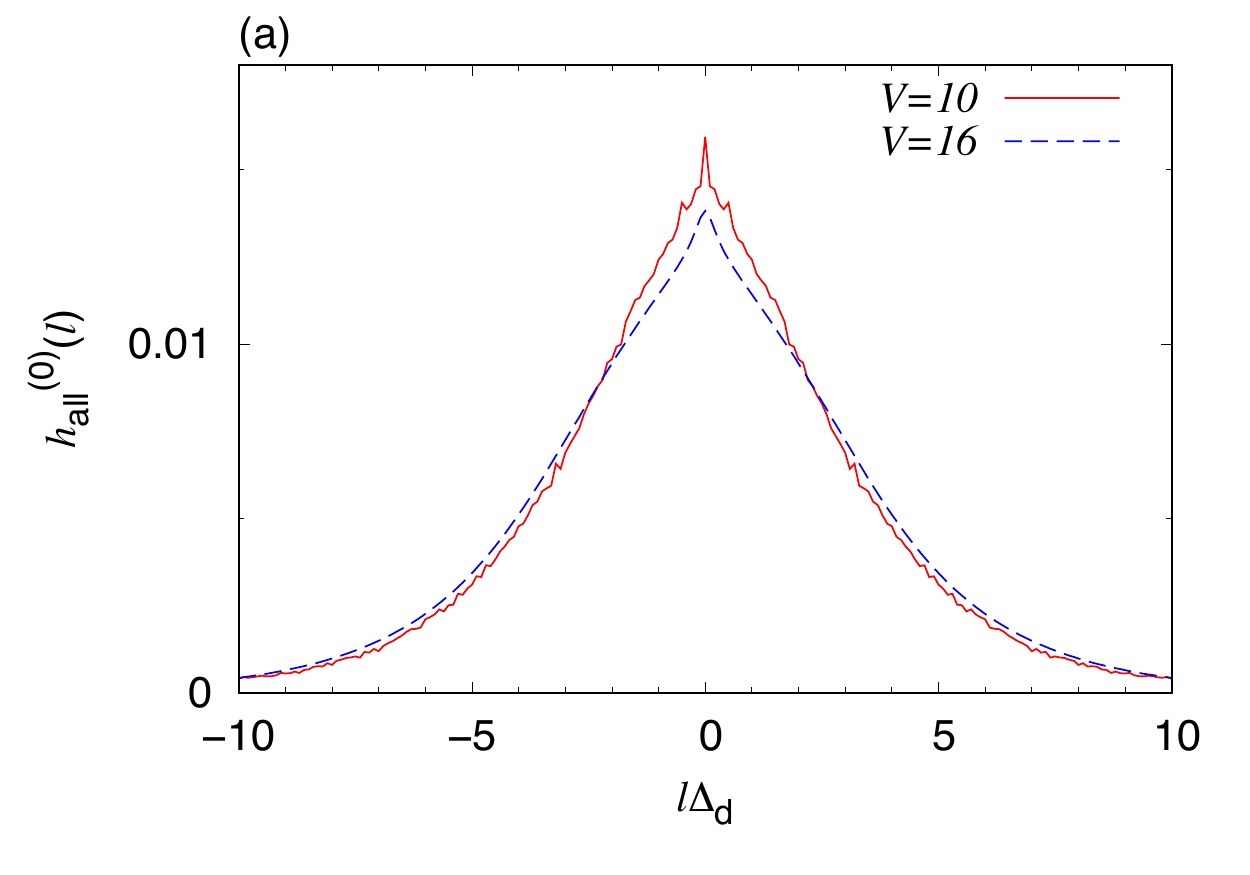}
\includegraphics[width=0.5\linewidth]{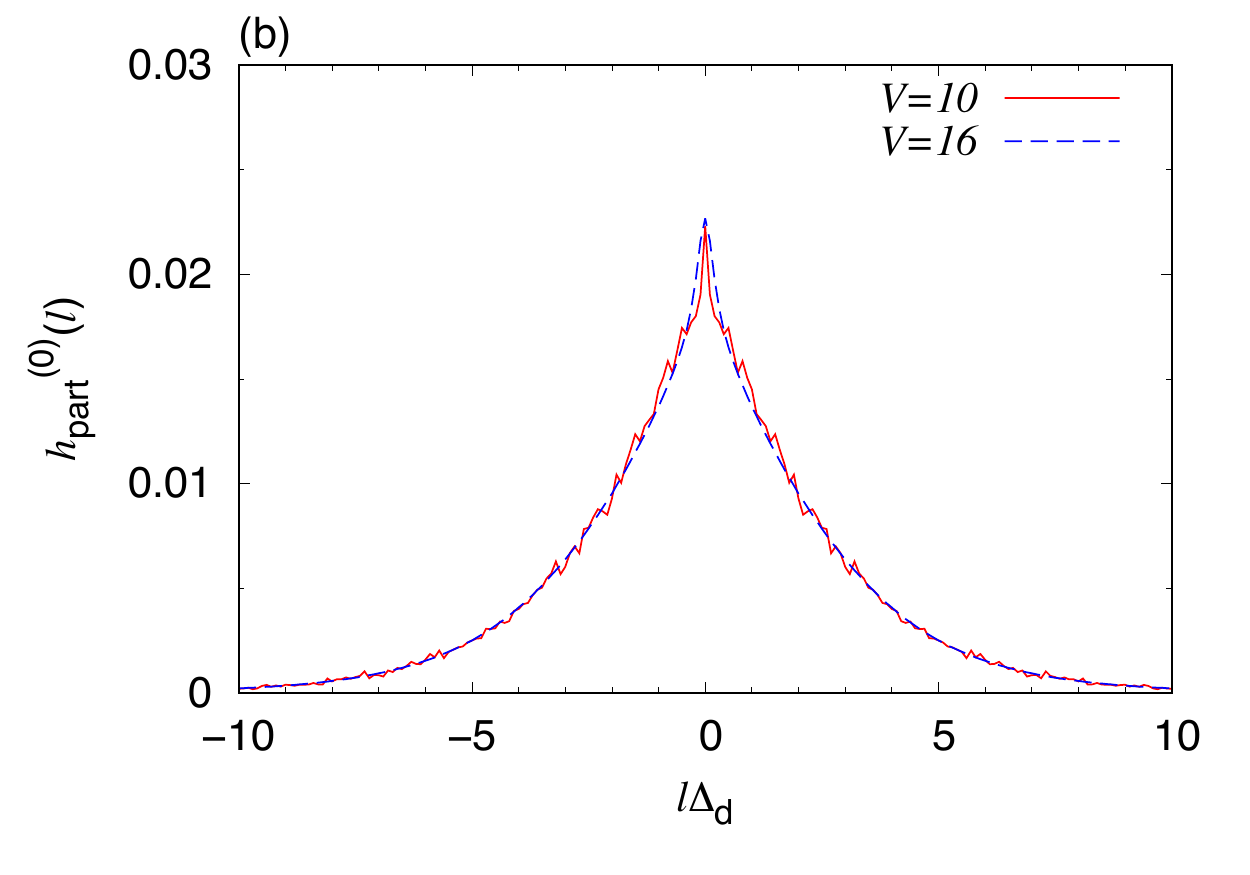}
\end{tabular}
\caption{
The scaled histograms for different system sizes, $V=10$ (red solid line) and $V=16$ (blue dotted line): (a) $h_{\rm all}^{(0)}(\ell)$ and (b) $h_{\rm part}^{(0)}(\ell)$.
}
\label{DB_v2}
\end{figure*}

In this appendix, we demonstrate the violation of the detailed balance condition in the model given by Eq.~(\ref{uniform}) in the main text.
The ``transition rate'' calculated by applying the perturbation theory is given by
\begin{equation}
W_{nm}=\sum_a|\braket{n|L_a|m}|^2,
\end{equation}
and the detailed balance condition with respect to the Gibbs state is expressed by
$\frac{W_{nm}}{W_{mn}}=e^{-\beta(E_n-E_m)}$
or equivalently,
\begin{equation}
\ln\frac{W_{nm}}{W_{mn}}+\beta(E_n-E_m)=0
\end{equation}
for any pair of eigenstates $n$ and $m$.

In order to judge whether the detailed balance condition holds in the model given by Eq.~(\ref{uniform}) in the main text, we make a scaled histogram for all the pairs of $n$ and $m$,
\begin{align}
h_{\rm all}(\ell)=&\frac{1}{2^V(2^V-1)} \sum_{\substack{n,m\\(n\neq m)}} \nonumber\\
&\times \int_{(\ell-1/2)\Delta_{\rm d}}^{(\ell+1/2)\Delta_{\rm d}} \delta \left( y-\ln\frac{W_{nm}}{W_{mn}} -\beta (E_n-E_m)\right)dy,
\end{align}
where $\ell\in \mathbb{Z}$ and $\Delta_{\rm d}=0.1$ is the bin size of the histogram and $\beta=0.256$ is the inverse effective temperature [see Appendix~\ref{app:temperature}].
The detailed balance condition holds when $h_{\rm all}(\ell)=\delta_{\ell,0}$.
In Fig.~\ref{DB}(a), we plot the scaled histogram for different system sizes, $V=10$ (red solid line) and $V=16$ (blue dotted line).
They are almost overlapped with each other, and have a peak at $\ell=0$ with finite width.
Thus, the detailed balance condition is violated irrespective of the system size.

In the histogram $h_{\rm all}(\ell)$, all the transition rates between the energy eigenstates are taken into account, but transition rates between the states with macroscopically different energies will be irrelevant to determine the steady state.
Thus, we produce another scaled histogram that omit such irrelevant contributions:
\begin{align}
h_{\rm part}(\ell)=&\frac{1}{\cal N}\Big[\sum_{\substack{n\\ \bar{E}-\Delta E\leq E_n \leq \bar{E}+\Delta E}} \Big[\sum_{\substack{m (m\neq n)\\ \bar{E}-\Delta E \leq E_m \leq \bar{E}+\Delta E}}\nonumber\\
&\times \int_{(\ell-1/2)\Delta_{\rm d}}^{(\ell+1/2)\Delta_{\rm d}} \delta \left( y-\ln\frac{W_{nm}}{W_{mn}} -\beta (E_n-E_m)\right)dy \Big]\Big],
\end{align}
where $\ell\in \mathbb{Z}$, $\bar{E}={\rm Tr} H \rho_{\rm s}^{(0)}$, and $\Delta E=4\simeq 1/\beta$.
Normalization constant ${\cal N}$ is determined from the condition that $\sum_\ell h_{\rm part}(\ell)=1$.
In Fig.~\ref{DB}(b), we plot the scaled histogram $h_{\rm part}(\ell)$ for different system sizes.
As in Fig.~\ref{DB}(a), it implies the violation of the detailed balance condition irrespective of the system size.

We also study the detailed balance condition with respect to $\{ P_n \}$ in Eq.~(\ref{eq:rho_0}).
We make the corresponding scaled histograms,
\begin{equation}
\left\{
\begin{aligned}
h_{\rm all}^{(0)}(\ell)=&\frac{1}{2^V(2^V-1)}\sum_{\substack{n,m\\(n\neq m)}} \int_{(\ell-1/2)\Delta_{\rm d}}^{(\ell+1/2)\Delta_{\rm d}} \delta \left( y-\ln\frac{W_{nm}P_m}{W_{mn}P_n}\right)dy,\\
h_{\rm part}^{(0)}(\ell)=&\frac{1}{{\cal N}^{(0)}}\Big[\sum_{\substack{n\\ \bar{E}-\Delta E\leq E_n \leq \bar{E}+\Delta E}} \Big[\sum_{\substack{m (m\neq n)\\ \bar{E}-\Delta E \leq E_m \leq \bar{E}+\Delta E}} \nonumber\\
& \qquad \times \int_{(\ell-1/2)\Delta_{\rm d}}^{(\ell+1/2)\Delta_{\rm d}} \delta \left( y-\ln\frac{W_{nm}P_m}{W_{mn}P_n} \right)dy \Big]\Big],
\end{aligned}
\right.
\end{equation}
where $\ell\in \mathbb{Z}$ and ${\cal N}^{(0)}$ is the normalization constant determined by $\sum_\ell h_{\rm part}^{(0)}(\ell)=1$.
In Figs.~\ref{DB_v2}, we plot the scaled histograms $h_{\rm all}^{(0)}(\ell)$ and $h_{\rm part}^{(0)}(\ell)$ for different system sizes,
which also imply the violation of the detailed balance condition with respect to $\{ P_n \}$.

\section{Emergence of the Gibbs state in another dissipative spin system}\label{app:another}
\begin{figure*}[t]
\centering
\includegraphics[width=0.49\linewidth]{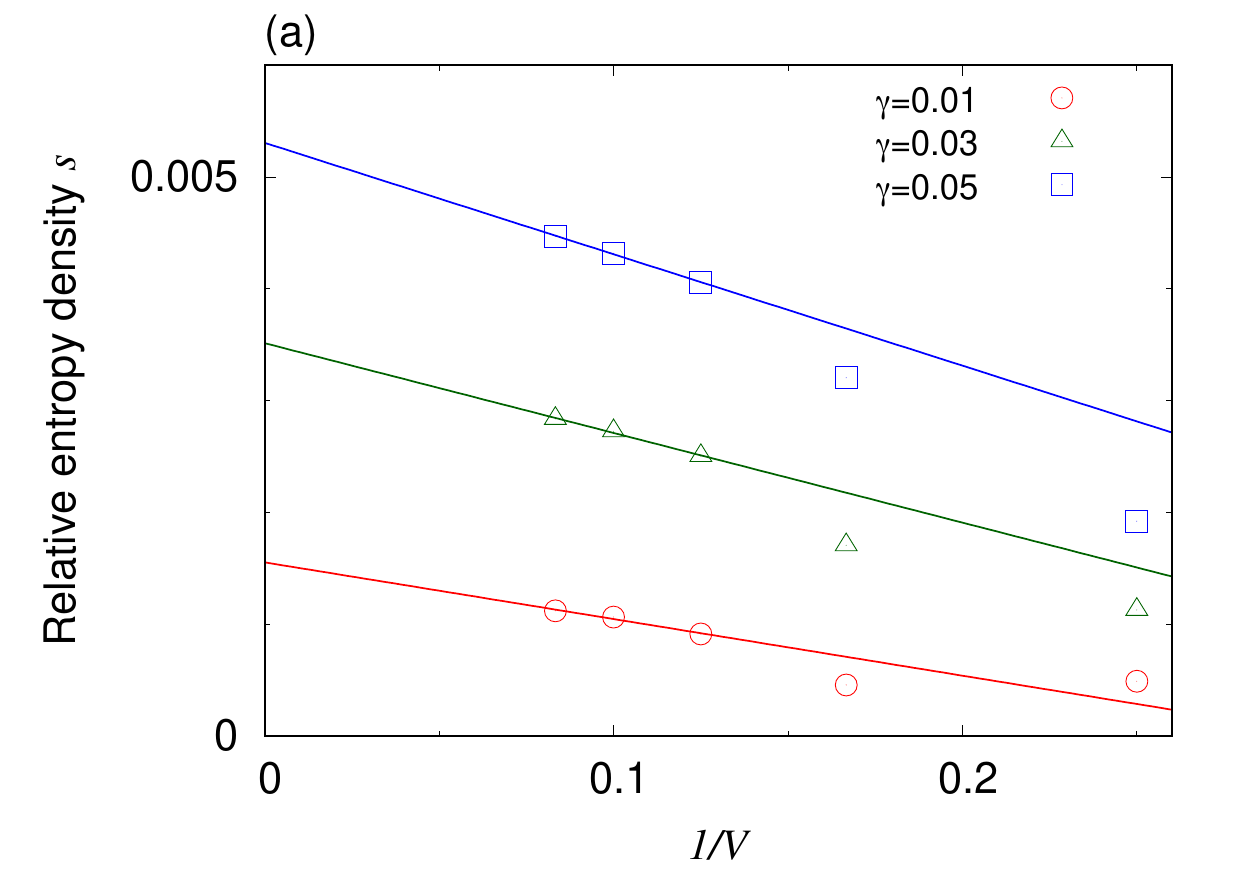}
\includegraphics[width=0.49\linewidth]{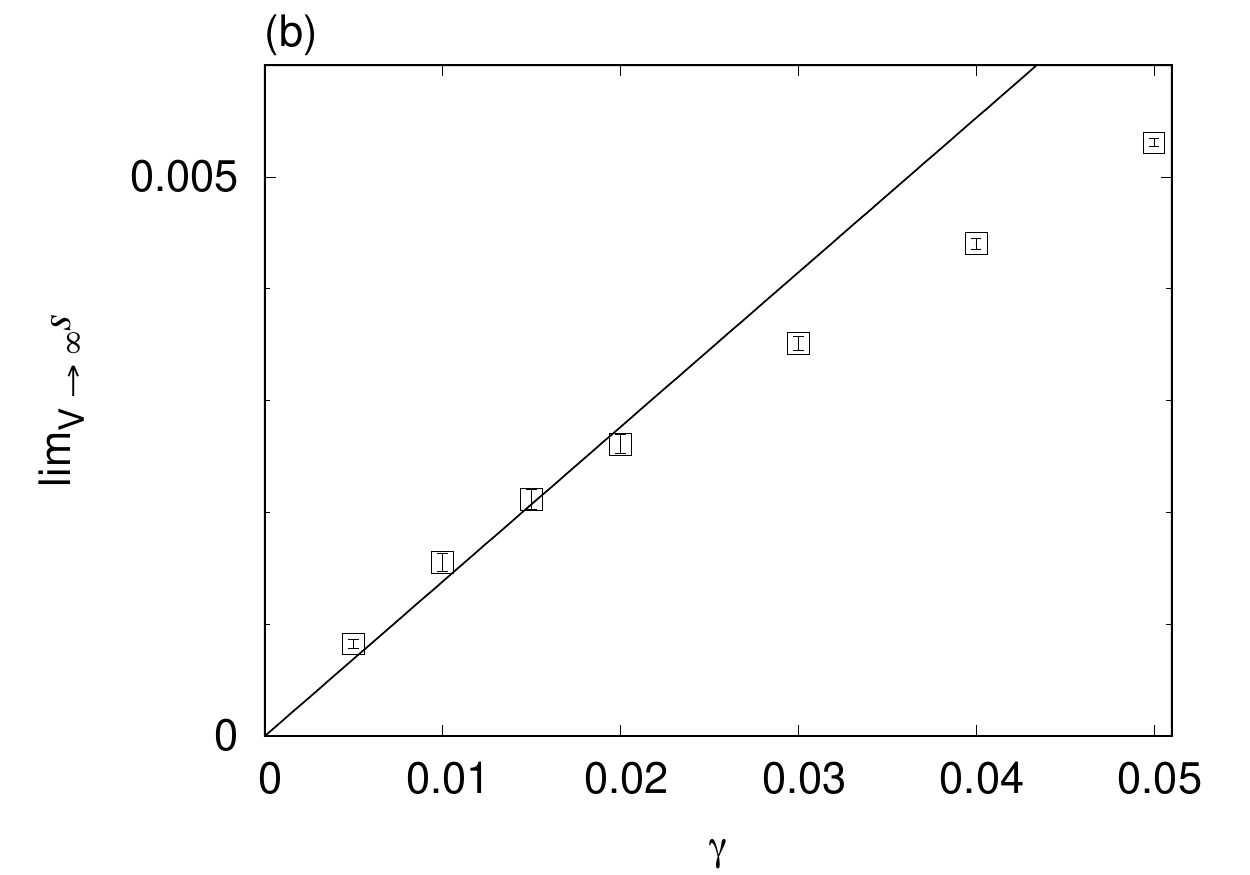}
\caption{
(a) System-size dependence of $s$ for the dissipative model [Eq.~(\ref{alternate})], in which no stationary current exists in the steady state.
Points are numerical data: $\gamma=0.01$(circle), $0.03$(triangle), and $0.05$(square).
The data for each $\gamma$ shows linear dependence on $1/V$ for large $V (V \ge 8)$.
(b) $\gamma$-dependence of $\lim_{V\to\infty}s$.
The full curve is a guide to show the dependence for small $\gamma$ $(\gamma \le 0.02)$, $\lim_{V\to\infty}s \propto \gamma$.
}
\label{size_alt}
\end{figure*}

We provide another model with bulk dissipation that shows the same $\gamma$-dependence of $\lim_{V \to \infty} s$ as the model in the main text. 
The Hamiltonian and the Lindblad operators are given by
\begin{equation}
\left\{
\begin{aligned}
H&=\sum_{i=1}^V (h^z S_i^z + h^x S_i^x) +\sum_{i=1}^{V-1} g S_i^z S_{i+1}^z,\\
L_i^{(1)}&= S_i^- \text{ for } i=1, 2,\cdots, V,\\
L_i^{(2)}&= S_i^+ \text{ for } i=2, 4,\cdots, V.
\end{aligned}
\right.
\label{alternate}
\end{equation}
where $(h^z, h^x, g)=(1.809, 1.618, 4)$.
We assume that $V$ is even.
There are two types of Lindblad operators: $L_i^{(1)}$ acting on every site and $L_i^{(2)}$ acting on even sites.
The system does not have any conserved current in the steady state, which implies that the steady state is described by a Gibbs state.

In Fig.~\ref{size_alt} (a), we show the system-size dependences of $s$ at $\gamma=0.01$, $0.03$, and $0.05$.
We find the linear dependence of the distance on $1/V$ for large $V$ ($V \ge 8$), and again we extrapolate the data for each $\gamma$ to the thermodynamic limit.
In Fig.~\ref{size_alt} (b), we present $\lim_{V\to\infty} s$ as a function of $\gamma$.
The figure shows $\lim_{V \to \infty} s \propto \gamma$, which is the same dependence as the model of the main text.


%

\end{document}